\documentclass[pdflatex,sn-basic]{sn-jnl}

\usepackage{graphicx}%
\usepackage{multirow}%
\usepackage{amsmath,amssymb,amsfonts}%
\usepackage{amsthm,amsfonts,amsmath,amssymb,bm}%
\usepackage{mathrsfs}%
\usepackage[title]{appendix}%
\usepackage{xcolor}%
\usepackage{textcomp}%
\usepackage{manyfoot}%
\usepackage{booktabs}%
\usepackage{algorithm}%
\usepackage{algorithmicx}%
\usepackage{algpseudocode}%
\usepackage{listings}%
\usepackage{ulem}%

\theoremstyle{thmstyleone}%
\theoremstyle{thmstyletwo}%

\theoremstyle{thmstylethree}%

\newcommand{\hi}{\textrm{H\textsc{i}}}
\newcommand{\fNL}{$f_{\rm NL}$}
\newcommand{\secref}[1]{\hyperref[#1]{Section~\ref*{#1}}}
\newcommand{\appref}[1]{\hyperref[#1]{Appendix~\ref*{#1}}}
\newcommand{\txteq}[1]{\,{#1}\,}
\newcommand{\Nfg}{N_\text{fg}}
\newcommand{\hMpc}{\,h\,\text{Mpc}^{-1}}
\newcommand{\degsq}{\,{\rm deg}^2}

\begin{document}

\title[Revealing cosmological fluctuations in 21cm intensity maps with MeerKLASS: from maps to power spectra]{Revealing cosmological fluctuations in 21cm intensity maps with MeerKLASS: from maps to power spectra}



\author*[1,2]{\fnm{Steven} \sur{Cunnington}}\email{steve.cunnington@port.ac.uk}

\author[3,4,5,6]{\fnm{Matilde} \sur{Barberi-Squarotti}}

\author[7]{\fnm{Jos\'e Luis} \sur{Bernal}}

\author[6,8,9,10]{\fnm{Stefano} \sur{Camera}}

\author[11,12]{\fnm{Isabella P.} \sur{Carucci}}

\author[13]{\fnm{Zhaoting} \sur{Chen}}

\author[14,15,10]{\fnm{Jos\'e} \sur{Fonseca}}

\author[10,16]{\fnm{Mario G.} \sur{Santos}}

\author[17,10]{\fnm{Marta} \sur{Spinelli}}

\author[18,10]{\fnm{Jingying} \sur{Wang}}

\author[2]{\fnm{Laura} \sur{Wolz}}

\affil[1]{\orgdiv{Institute of Cosmology \& Gravitation}, \orgname{University of Portsmouth}, \orgaddress{\street{Dennis Sciama Building, Burnaby Road}, \city{Portsmouth}, \postcode{PO1 3FX}, \country{UK}}}

\affil[2]{\orgdiv{Jodrell Bank Centre for Astrophysics, Department of Physics \& Astronomy}, \orgname{The University of Manchester}, \city{Manchester}, \postcode{M13 9PL}, \country{UK}}

\affil[3]{\orgdiv{Dipartimento di Fisica, Universit\`a degli Studi di Milano, via G.\ Celoria 16, 20133 Milano, Italy}}

\affil[4]{\orgdiv{INFN – Istituto Nazionale di Fisica Nucleare, Sezione di Milano, via G.\ Celoria 16, 20133 Milano, Italy}}

\affil[5]{\orgdiv{INAF -- Istituto Nazionale di Astrofisica, Osservatorio Astrofisico di Brera-Merate, via Brera 28, 20121 Milano, Italy}}

\affil[6]{\orgdiv{Dipartimento di Fisica, Universit\`a degli Studi di Torino, Via P.\ Giuria 1, 10125 Torino, Italy}}

\affil[7]{\orgdiv{Instituto de F{\'i}sica de Cantabria (IFCA)}, \orgname{CSIC-Univ. de Cantabria}, \orgaddress{\street{Avda. de los Castros s/n,}, \city{Santander}, \postcode{E-39005}, \country{Spain}}}

\affil[8]{\orgdiv{INFN -- Istituto Nazionale di Fisica Nucleare, Sezione di Torino, Via P.\ Giuria 1, 10125 Torino, Italy}}

\affil[9]{\orgdiv{INAF -- Istituto Nazionale di Astrofisica, Osservatorio Astrofisico di Torino, Strada Osservatorio 20, 10025 Pino Torinese, Italy}}

\affil[10]{\orgdiv{Department of Physics \& Astronomy, University of the Western Cape, Cape Town 7535, South Africa}}

\affil[11]{\orgdiv{INAF -- Istituto Nazionale di Astrofisica}, \orgname{Osservatorio Astronomico di Trieste}, \orgaddress{\street{Via G.B.\ Tiepolo 11}, \city{Trieste}, \postcode{34131}, \country{Italy}}}

\affil[12]{\orgdiv{IFPU -- Institute for Fundamental Physics of the Universe}, \orgaddress{\street{Via Beirut 2}, \city{Trieste}, \postcode{34151}, \country{Italy}}}

\affil[13]{\orgdiv{Institute for Astronomy, Royal Observatory}, \orgname{The University of Edinburgh}, \city{Edinburgh}, \postcode{EH9 3HJ}, \country{UK}}

\affil[14]{\orgdiv{Instituto de Astrof\'isica e Ci\^encias do Espa\c{c}o}, \orgname{Universidade do Porto}, \orgaddress{CAUP, Porto \street{Rua das Estrelas}, \city{Porto}, \postcode{4150-762}, \country{Portugal}}}

\affil[15]{\orgdiv{Departamento de F\'isica e Astronomia, Faculdade de Ci\^{e}ncias}, \orgname{Universidade do Porto}, \orgaddress{CAUP, Porto \street{Rua do Campo Alegre 687}, \city{Porto}, \postcode{4169-007}, \country{Portugal}}}

\affil[16]{\orgdiv{South African Radio Astronomy Observatory (SARAO), Liesbeek House, River Park, Gloucester Road, Mowbray, Cape Town, 7700, South Africa}}

\affil[17]{\orgdiv{Observatoire de la C\^{o}te d’Azur, Laboratoire Lagrange, Bd de l’Observatoire, CS 34229, 06304 Nice cedex 4, France}}

\affil[18]{\orgdiv{State Key Laboratory of Radio Astronomy and Technology, Shanghai Astronomical Observatory, CAS, 80 Nandan Road, Shanghai 200030, China}}

\abstract{Mapping the integrated 21cm emission line from dark matter-tracing neutral hydrogen gas is the primary science goal for MeerKLASS (MeerKAT's Large Area Synoptic Survey). Prior to the arrival of MeerKAT, this intensity mapping technique had only been tested on a couple of pre-existing single-dish radio telescopes with a handful of observational hours with which to make early pioneering detections. The 64-dish MeerKAT array, precursor to the SKA Observatory (SKAO), can scan the sky in auto-correlation \textcolor{black}{(or \textit{single-dish})} mode and perform intensity mapping across large sky areas, presenting the exciting potential for a wide area (${\gtrsim}\,10{,}000\,{\rm deg}^2$) spectroscopic survey across redshift $0.4\txteq{<}z\txteq{<}1.45$. Validating the \textcolor{black}{single-dish} mode of observation for a multi-dish array and developing the analysis pipeline with which to make unbiased measurements has presented major challenges to this endeavour. In this work, we overview the advances in the field that have facilitated a robust analysis framework for single-dish intensity mapping, and review some results that showcase its success using early MeerKLASS surveys. We demonstrate our control of foreground cleaning, signal loss and map regridding to deliver detections of cosmological clustering within the intensity maps through cross-correlation power spectrum measurements with overlapping galaxy surveys. Finally, we discuss the prospects for future MeerKLASS observations and forecast its potential, making our code publicly available: \href{https://github.com/meerklass/MeerFish}{\texttt{https://github.com/meerklass/MeerFish}}.}




\maketitle

\section{Introduction}\label{sec1}

Mapping the large-scale structure of the Universe through the redshifted 21cm emission line of neutral hydrogen (\hi) is emerging as a powerful technique in observational cosmology. By surveying the diffuse \hi\ signal without resolving individual galaxies, \hi\ intensity mapping offers a means to efficiently chart matter fluctuations across vast cosmological areas and redshift ranges \citep{Bharadwaj:2000av,Battye:2004re,Wyithe:2007rq,Chang:2007xk,Liu:2019awk,Villaescusa-Navarro:2018vsg}.

Prior to the arrival of MeerKAT\footnote{\href{https://www.sarao.ac.za/science/meerkat/}{sarao.ac.za/science/meerkat/}}, CHIME \citep{CHIME:2022dwe}, FAST \citep{Hu:2019okh} and uGMRT \citep{Elahi:2024qbp}, 21cm intensity mapping had only been explored through a handful of pioneering efforts using existing single-dish radio telescopes such as the Green Bank Telescope (GBT) and Parkes, with restricted sky coverage and observational time \citep{Masui:2012zc,Switzer:2013ewa,Wolz:2015lwa,Anderson:2017ert,eBOSS:2021ebm}. These early detections demonstrated the feasibility of cross-correlation analyses between intensity maps and galaxy surveys, but enhanced sensitivity and a vast increase in observational resources would be needed to achieve competitive cosmological constraints. 

\textcolor{black}{The advent of MeerKAT, a 64-dish precursor to the SKA Observatory (SKAO)\footnote{\href{https://www.skao.int/en}{skao.int/en}}, has opened a new chapter in this field and will be the focus of this paper. However, other exciting experiments are also underway. CHIME has reported detections of the 21cm emission in cross-correlation with the eBOSS ELG, LRG and QSO-galaxies \citep{CHIME:2022kvg} and the Ly$\alpha$ forest at $z\txteq{\sim}2.3$ \citep{CHIME:2023til}. Plus more recently, an auto-power spectrum detection of the cosmological 21cm signal on ${\sim}\,1$\,Mpc scales \citep{CHIME:2025cee} was claimed. uGMRT has delivered competitive upper limits on the 21cm power spectrum at $z\txteq{\sim}2.3$ through tapered gridded estimator analysis \citep{Elahi:2024qbp}. Meanwhile, FAST is also demonstrating the feasibility of 21cm intensity mapping, delivering validated calibration pipelines, stable noise performance, and foreground-cleaned maps over hundreds of square degrees using early drift-scan data \citep{Li:2023zer,Yang:2024hos}. More intensity mapping projects are also planned and are expected to begin contributions soon \citep{Vanderlinde:2019tjt,Crichton:2021hlc,Abdalla:2021nyj}.} 

MeerKAT's Large Area Synoptic Survey (MeerKLASS)\footnote{\href{https://meerklass.org/}{meerklass.org/}} \citep{MeerKLASS:2017vgf} is a major effort to realise the scientific potential of \hi\ intensity mapping in the pre-SKAO era. While the primary operation of the MeerKAT array is as an \textit{interferometer}, the distribution of its baselines is not well-suited to the very large angular scales required for \hi\ intensity mapping on ``cosmological'' scales (${\lesssim}\,0.2\hMpc$) \citep{Bull:2014rha}. The solution is instead to use the auto-correlation of the voltage signal from each of the 64 dishes, effectively operating the array in \textit{single-dish} mode\footnote{\textcolor{black}{From hereon we will exclusively refer to the \textit{dish} auto-correlation mode of observations as ``single-dish mode'' to help distinguish it from the different concept of auto-correlation (\hi-\hi) \textit{power spectrum} measurements, discussed later.}} to gather wide-field maps of the sky for each individual dish. This approach allows MeerKLASS to target the detection of the 21cm signal across thousands of square degrees, while at the same time retaining the interferometric visibilities for a wide range of commensal science. The frequency coverage from using MeerKAT's UHF-band receiver will allow access to redshifts up to $z\txteq{=}1.45$ \citep[SKAO Band 1 will extend this to $z\txteq{=}3$ ][]{SKA:2018ckk}, making single-dish 21cm intensity mapping with these instruments ideally suited for vast volume spectroscopic surveys.

Validating the single-dish mode of observation on a multi-dish array primarily designed for interferometry has presented significant technical and methodological challenges. Unlike traditional interferometric experiments, single-dish intensity mapping with MeerKAT has required a dedicated pipeline to calibrate, map-make, clean, and statistically analyse large volumes of novel data. Dominant foregrounds \citep{Santos:2004ju,Liu:2009qga,Wolz:2013wna,Alonso:2014sna}, radio frequency interference (RFI) \citep{Harper:2018ncl,2019JAI.....840010B,Engelbrecht:2024eoc}, correlated noise \citep{Li:2020bcr,Irfan:2023njr}, beam systematics \citep{2021MNRAS.502.2970A,Matshawule:2020fjz,Spinelli:2021emp,Chen:2025bcx}, and the inherently faint amplitude of the \hi\ brightness temperature, all compound the difficulty of extracting unbiased cosmological measurements.

To meet these challenges, the MeerKLASS collaboration has developed a robust data treatment and analysis framework. Iterative self-calibration has been explored \citep{Wang:2020lkn,MeerKLASS:2024ypg} to suppress model reliance. Foreground cleaning methods, particularly blind component separation techniques \citep{Alonso:2014dhk,Carucci:2020enz,Cunnington:2020njn,Spinelli:2021emp} such as Principal Component Analysis (PCA), are employed to reduce astrophysical foregrounds. Carefully accounting for signal loss from foreground cleaning is also crucial, and currently quantified and corrected using foreground transfer functions derived through mock signal injection into the real data \citep{Switzer:2015ria,Cunnington:2023jpq}. Another critical development has been the regridding of spherical sky maps into Cartesian volumes suitable for Fourier analysis, avoiding significant biases in large-sky surveys \citep{Cunnington:2023aou}. With these tools in hand, MeerKLASS has achieved statistically significant detections of \hi\ clustering through cross-correlations with overlapping spectroscopic galaxy surveys such as WiggleZ and GAMA \citep{Cunnington:2022uzo,MeerKLASS:2024ypg,Carucci:2024qpm}. These detections have marked a key milestone: the first robust demonstration of single-dish intensity mapping with a multi-dish array, laying the groundwork for full-scale cosmological analyses with the extended MeerKLASS and the upcoming SKAO.

This paper provides an overview of these advances, summarising the core elements of the MeerKLASS analysis pipeline and highlighting the methods used to overcome the major systematic challenges facing \hi\ intensity mapping. In \secref{sec:Observations}, we describe the MeerKLASS observational strategy and data. \secref{sec:FGcleaning} presents the approaches to foreground cleaning and signal loss correction, including the role of the foreground transfer function (\secref{sec:TF}). \secref{sec:Regridding} discusses the regridding of sky maps into Cartesian volumes for power spectrum estimation. In \secref{sec:SignalDetections}, we present the cross-correlation power spectrum measurements and their implications for cosmological constraints. In \secref{sec:Future} we look to future prospects, and forecast the potential of the full MeerKLASS, before summarising in \secref{sec:Summary}.

\section{Observations with MeerKLASS}\label{sec:Observations}

Since 2018, MeerKLASS has been using the MeerKAT array \citep{2016mks..confE...1J} to steadily build survey area and sensitivity with single-dish mode observations. Initially, all observations were done using MeerKAT's L-band, which has a receiver bandpass of $900\txteq{<}\nu\txteq{<}1670\,$MHz. However, since large cosmological volumes only become available for redshifted-21cm at ${\lesssim}\,1200\,$MHz, plus RFI contamination further limits this window to $971\txteq{<}\nu\txteq{<}1023\,$MHz, the useable cosmological redshift coverage is limited to $0.39\txteq{<}z\txteq{<}0.46$. This partly motivated the move to the lower frequency receiver in UHF-band, which we discuss shortly. The first pilot MeerKLASS L-band survey was completed in 2019 \citep{Wang:2020lkn}, and then a deep-field survey in late 2021 \citep{MeerKLASS:2024ypg}. This data served as essential proving grounds for developing calibration, map-making and analysis pipelines, and led to the first cosmological clustering detections with MeerKAT \citep{Cunnington:2022uzo}, much of which we will review.

The current MeerKLASS approach is to conduct single-dish observations with a fast scanning strategy to suppress the impact of slow instrumental gain variations on the angular scales of interest for cosmology. The dishes scan back-and-forth in azimuth at constant elevation to minimise ground pickup and atmospheric fluctuations, with noise diodes firing at regular intervals to provide a stable reference for gain calibration. This approach ensures reasonably uniform sky coverage while maintaining instrumental stability. In parallel, the interferometric visibilities are also recorded in an on-the-fly (OTF) mode to support commensal science, and will be the subject of forthcoming publications \citep{ChatterjeeOTF,ManglaOTF,PaulOTF}. We refer those readers interested in the technical implementation of this pipeline to \citet{Wang:2020lkn}, where the calibration, RFI flagging and map-making methodology was pioneered for this unique data (also see \citet{Li:2020bcr,Irfan:2021xuk,MeerKLASS:2024ypg} for related developments). The overview in this paper will not go into detail on this aspect and instead focuses on the post-calibration, analysis workflow, i.e.\ \textit{from maps to power spectra}, as titled.

MeerKLASS is now entering a mature phase. With large observational campaigns planned, it presents a fantastic opportunity for precision cosmology at radio wavelengths and complementarity with upcoming galaxy surveys such as DESI, Euclid, 4MOST and the Rubin Observatory \citep{DESI:2016fyo,Euclid:2019clj,2019Msngr.175....3D,LSSTDarkEnergyScience:2018jkl}. \autoref{tab:MKobs} summarises the cumulative MeerKLASS observations, both completed and planned. MeerKLASS is now steering towards using the UHF-band receiver in the most recent and future observations. As discussed, we found that the L-band receiver's useful cosmological range was limited to $0.39\txteq{<}z\txteq{<}0.46$. The UHF-band however, offers a dramatically wider redshift window ($0.4\txteq{<}z\txteq{<}1.45$), which early observations are showing is more robust to RFI contamination. The move to UHF holds crucial gains for cosmology since the observed comoving volume will be 40 times larger than an equivalent area of observation in the L-band.

\begin{table}[!htb]
    \vspace{-5pt}
    \caption{MeerKLASS observations by year. Observing time is split into two columns:  yearly added and \textit{cumulative} total for each MeerKAT telescope receiver. Total area is also the amassed cumulative total for each receiver. Awarded and extension status (last column) is based on the latest MeerKAT extra-large proposal (XLP) outcomes. The final two rows demonstrate a hypothetical continuation of MeerKLASS, given MeerKAT remains a distinct instrument from the SKAO-array. All forecasts (shown later) only assume the extension XLP time up to 2028, i.e.\ 2{,}500 and $10{,}000\degsq$.}
    \begin{tabular}{ccccccc}    \vspace{-15pt}
\\\toprule
         & \multicolumn{2}{c}{\textbf{Obs. time}} & 
        \textbf{Total} &  &  & \\
        \textbf{Year} & (Added) & (Total) & \textbf{area} & \textbf{Receiver} & \textbf{Redshift}$\mathbf{^{*}}$ & \textbf{\textcolor{black}{Status as of 2026}} \\
        & [hrs] & [hrs] & [deg$^2$] & & &
        \\\midrule
        2019 & +15$^{**}$ & 15 & 250 & L & $0.39\txteq{<}z\txteq{<}0.46$ & Observed \\
        2021 & +70 & 85 & 500 & L & $0.39\txteq{<}z\txteq{<}0.46$ & Observed \\\midrule
        2023 & +110 & 110 & 500 & UHF & $0.4\txteq{<}z\txteq{<}1.45$ & Observed \\
        2024 & +270 & 380 & 1{,}600 & UHF & $0.4\txteq{<}z\txteq{<}1.45$ & Observed \\
        2025 & +500 & 880 & 3{,}600 & UHF & $0.4\txteq{<}z\txteq{<}1.45$ & Observing \\
        2026 & +500 & 1{,}380 & 5{,}600 & UHF & $0.4\txteq{<}z\txteq{<}1.45$ & Awarded under XLP \\
        2027 & +550 & 1{,}930 & 7{,}800 & UHF & $0.4\txteq{<}z\txteq{<}1.45$ & Extension under XLP \\
        2028 & +570 & 2{,}500 & 10{,}000 & UHF & $0.4\txteq{<}z\txteq{<}1.45$ & Extension under XLP \\\midrule
        2029 & +500 & 3{,}000 & 12{,}000 & UHF & $0.4\txteq{<}z\txteq{<}1.45$ & Pre-SKAO continuation \\
        2030 & +500 & 3{,}500 & 14{,}000 & UHF & $0.4\txteq{<}z\txteq{<}1.45$  & Pre-SKAO continuation \\
        \bottomrule     
    \end{tabular}
    \vspace{-0pt}
    {$\mathbf{^{*}}$ \textit{Useable} redshift is given. For L-band, this is severely restricted due to RFI. For UHF, there will also be some RFI restrictions, but this is more contained to narrow chunks of missing channels; thus, for simplicity, we list the full UHF range.\newline
    $^{**}$ Over 2018/19, 36 hours in total were observed, but several teething problems meant that only 15\,hrs were kept, and of this, only ${\sim}10$\,hrs were used in analysis (shown later).}
    \label{tab:MKobs}
    \vspace{-20pt}
\end{table}

To date, MeerKLASS has amassed over 650 hours of UHF data reaching nearly 3,000 deg$^2$, already making it the largest spectroscopic survey of large-scale structure in the Southern hemisphere. The near future will see an increase to 2,500 hours and 10,000 deg$^2$ coverage by 2028, with possible continuation beyond into the pre-SKAO era. In the final \secref{sec:Future}, before summarising, we will demonstrate the cutting-edge science this programme has the potential to deliver with this data stream in UHF.

\section{Foreground cleaning}\label{sec:FGcleaning}

The 21cm intensity mapping technique measures the integrated radio emission across large areas of the sky and over a wide range of frequencies. By design, it collects all contributions to the sky brightness temperature, not only the faint redshifted \hi\ signal but also much brighter \textit{contaminants}. Among these, the most challenging are astrophysical foregrounds, which are many orders of magnitude stronger than the cosmological signal of interest.

The dominant foreground component at the relevant frequencies is synchrotron emission, arising from relativistic electrons spiralling in Galactic magnetic fields. This diffuse emission is spectrally smooth but extremely bright, overwhelming the expected \hi\ fluctuations. A secondary, but still significant, contribution comes from Galactic free–free emission, produced by the scattering of electrons off ions in ionised regions. Both components present strong, smooth frequency spectra that contrast with the fluctuating nature of the cosmological \hi\ signal.

For an unbiased measurement of large-scale structure, these astrophysical foregrounds must be removed with high precision \citep{Ansari:2011bv}. Because detailed foreground models remain uncertain, most successful approaches in current analyses adopt ``blind'' cleaning strategies that rely on the spectral smoothness of foregrounds rather than prior knowledge of their exact structure and amplitude \citep{Alonso:2014dhk,Carucci:2020enz,Cunnington:2020njn,Spinelli:2021emp}. Principal Component Analysis (PCA)-based foreground cleaning has become one of the most widely used techniques \citep{Masui:2012zc,Anderson:2017ert,Cunnington:2022uzo,Carucci:2024qpm,MeerKLASS:2024ypg,Chen:2025bcx}, exploiting statistical decompositions of the data to identify and subtract dominant smooth modes while preserving the cosmological signal.

\subsection{Principal Component Analysis (PCA) and its extensions}\label{sec:PCA}

A common approach to foreground removal in \hi\ intensity mapping is to exploit the expected spectral smoothness of the contaminating astrophysical signals, compared to the more rapidly varying cosmological \hi\ fluctuations. The most widely applied \textit{blind} component separation method is Principal Component Analysis (PCA). For each frequency channel $\nu$ we define a map of the total observed brightness temperature $X_{\rm obs}(\nu, \boldsymbol{\theta})$ at sky pixel $p$ as
\begin{equation}
    X_{\rm obs}(\nu,\boldsymbol{\theta}) = T_{\mathrm{FG}}(\nu,\boldsymbol{\theta}) + T_{\hi}(\nu,\boldsymbol{\theta}) + T_{\mathrm{N}}(\nu,\boldsymbol{\theta}) \, ,
\end{equation}
where $T_{\mathrm{FG}}$ are the foreground contaminants, $T_{\hi}$ is the cosmological \hi\ signal, and $T_{\mathrm{N}}$ represents instrumental noise. In matrix form, the full data cube can be written as
\begin{equation}
    \textbf{\textsf{X}}_{\rm obs} = \textbf{\textsf{A}}\,\textbf{\textsf{A}}^{\top}\textbf{\textsf{X}}_{\rm obs} + \textbf{\textsf{R}} \, ,
\end{equation}
where $\textbf{\textsf{A}}$ is the $N_{\nu} \times N_{\mathrm{fg}}$ mixing matrix acting on the observations to isolate the contamination, which we want to remove, leaving the residual term $\textbf{\textsf{R}}$, containing cosmological signal and instrumental noise. The aim of the cleaning procedure is to estimate the contaminant contribution $\hat{\textbf{\textsf{A}}}\hat{\textbf{\textsf{S}}}$ 
and project it out of the data, giving
\begin{equation}
    \textbf{\textsf{X}}_{\mathrm{clean}} = \textbf{\textsf{X}}_{\rm obs} - \hat{\textbf{\textsf{A}}}\,  \hat{\textbf{\textsf{A}}}^{\top} \textbf{\textsf{X}}_{\rm obs} \, .
\end{equation}
PCA provides an estimate of $\hat{\textbf{\textsf{A}}}$ by identifying the set of eigenvectors \textbf{\textsf{V}} of the frequency--frequency covariance matrix with the largest eigenvalues. These modes capture the majority of the variance in the data, which is assumed to be dominated by spectrally-smooth foregrounds. The first $N_{\mathrm{fg}}$ components are therefore removed, while the remaining modes are retained as the cleaned data. The choice of $N_{\mathrm{fg}}$ is not fixed a priori, and is a key challenge for an optimal PCA-based foreground cleaning, since low $\Nfg$ increases residual contamination but too high $\Nfg$ causes cosmological signal loss, which we discuss further shortly.

\subsubsection{Multiscale extension (mPCA)}\label{sec:mPCA}

In reality, instrumental and environmental systematics interact and mix with the foregrounds, partially corrupting their spectral smoothness and rendering the cleaning problem more challenging. In particular, given the enormous discrepancy in intensity between the foregrounds and the  \hi\ signal, any tiny residual of uncharacterised contaminant becomes destructive for the extraction of the signal. 

Instead of relying only on the radial characterisation of the contaminants, it is also helpful to consider the pixel-domain (angular) information. The rationale is that the systematic influence over astrophysical foreground happens differently at different spatial angular scales, being more prominent at the small scales. Instead, at large scales, the astrophysical contribution remains the most dominant. We have put this idea into practice with MeerKLASS data \citep{Carucci:2024qpm}, designing a multiscale PCA (mPCA) algorithm that we compared to standard PCA. 

In this multiscale approach, each frequency map is first decomposed into wavelet-filtered large- and small-scale maps, using the isotropic undecimated wavelet transform \citep{Starck2007}. From one data cube, we now have two cubes of the same size (each of them referring to the large- and small-scale only information) which are then cleaned separately: the contaminant separation problem becomes two independent problems, with their own, independently defined mixing matrix and number of removed components ($N_{\mathrm{fg}}$). This allows for a better characterisation of the two contaminant `regimes'. The final cleaned map is the sum of the independently cleaned small- and large-scale maps, since the wavelet decomposition is exact. 

By using one of our cross-correlation detections as a benchmark \citep{Cunnington:2022uzo}, we could test the performance of mPCA. We found mPCA led to a higher and more scale-independent cross-correlation signal, plus evidence for less residual contamination. The mPCA maps also had less prominent oﬀ-diagonal terms in its frequency-correlation matrix, suggesting less mode-mixing between residual contaminants and signal, 
making it more robust. We refer to \citet{Carucci:2024qpm} for more details. In summary, mPCA offers a promising extension for future \hi\ intensity mapping surveys.

\subsection{Signal loss correction: the foreground transfer function}\label{sec:TF}

Blind foreground cleaning has shown impressive progress in removing the overwhelming astrophysical contaminants, enabling the recovery of the cosmological \hi\ brightness fluctuations. However, since the cleaning process projects out correlated modes along the line of sight, a fraction of the true \hi\ signal will inevitably be removed as well \citep{Switzer:2015ria,Cheng:2018osq}. This effect, known as \textit{signal loss}, is unavoidable and must be carefully corrected for. 

Whilst considerable effort has been devoted to mitigating the bias from foreground \textit{residuals}, the equally important problem of \textit{over-cleaning} is less often addressed. Over-cleaning leads to signal loss that can substantially suppress the measured power spectrum, particularly on large radial scales most degenerate with foregrounds. Unlike residual contamination, the bias from signal loss is not mitigated by cross-correlations, thus arguably presenting a more threatening problem. The MeerKLASS collaboration has therefore carried out dedicated studies of signal loss, with the aim of ensuring robust cosmological interpretation of foreground-cleaned maps \citep{Cunnington:2023jpq}.

A standard approach to correct for signal loss is through the construction of a foreground transfer function, $\mathcal{T}(\boldsymbol{k})$, which quantifies the attenuation of the true power spectrum due to the cleaning process. The transfer function is defined such that
\begin{equation}
    P_{\rm clean}(\boldsymbol{k}) = \mathcal{T}(\boldsymbol{k}) P_{\rm true}(\boldsymbol{k}) \, .
\end{equation}
Thus, a robust process to estimate the transfer function, $\widehat{\mathcal{T}}(\boldsymbol{k})$, is required to unbias (or reconstruct) the signal at the power spectrum level. The reconstructed power spectrum, on which modelling and cosmological inference is performed, is then simply given as
\begin{equation}
    P_{\rm rec}(\boldsymbol{k}) \;=\; \frac{P_{\rm clean}(\boldsymbol{k})}{\widehat{\mathcal{T}}(\boldsymbol{k})} \, .
\end{equation}
In practice, $\widehat{\mathcal{T}}(\boldsymbol{k})$ is calculated pseudo-empirically by injecting mock \hi\ signals into the \textit{real} contaminated data, applying the same cleaning procedure, and measuring the fractional suppression of the mock power spectrum \citep{Switzer:2015ria}. The reason for using the actual data itself is that accurate emulation of the foregrounds and their response to the complex instrumentation is extremely challenging and would inevitably result in underestimation of the contamination.

\textcolor{black}{The transfer function correction is only empirically accessible through mock signal injection, meaning its accuracy ultimately relies on the assumption that the injected mocks adequately sample the relevant signal–foreground couplings present in the real data. Large numbers of mock simulations ${\gtrsim}\,100$ are typically required to reach a converged transfer function, making this a computationally expensive process. Furthermore, as recently clarified by \citep{Chen:2025til}, blind cleaning induces mode-mixing in Fourier space, which will propagate into the transfer function calculation. We discuss some of the consequences of this at the end of this section.}

\medskip
\noindent
\textbf{Construction recipe}: Following \citet{Cunnington:2023jpq}, the recipe for building an unbiased transfer function proceeds as:
\begin{enumerate}
    \item Foreground clean the observed maps,
    \begin{equation}
        \textbf{\textsf{X}}_{\rm clean} = \textbf{\textsf{X}}_{\rm obs} - \textbf{\textsf{V}}\, \textbf{\textsf{S}}\, \textbf{\textsf{V}}^{\top}\, \textbf{\textsf{X}}_{\rm obs} \, ,
    \end{equation}
    where $\textbf{\textsf{X}}_{\rm obs}$ are the observed maps, $\textbf{\textsf{V}}$ the matrix of eigenmodes from the frequency–frequency covariance, and $\textbf{\textsf{S}}$ a 
    diagonal selection matrix specifying the $N_{\rm fg}$ modes to remove.
    \item Generate mock \hi\ realisations with the same 
    geometry as the data. We indicate these mocks by $\textbf{\textsf{M}}$ to clearly distinguish them from real data $\textbf{\textsf{X}}$.
    \item Inject the mock into the data and repeat the foreground clean, 
    subtracting the cleaned data without injection to suppress variance, giving the cleaned mock map
    \begin{equation}\label{eq:Mclean}
        \textbf{\textsf{M}}_{\rm clean} = (\textbf{\textsf{X}}_{\rm obs} + \textbf{\textsf{M}})
        - \textbf{\textsf{V}}_{\rm obs+m}\, \textbf{\textsf{S}}\, \textbf{\textsf{V}}_{\rm obs+m}^{\top}\, (\textbf{\textsf{X}}_{\rm obs} + \textbf{\textsf{M}})
        - [\textbf{\textsf{X}}_{\rm clean}] \, .
    \end{equation}
    \item Estimate the transfer function as the ratio of cross- and auto-spectra:
    \begin{equation}\label{eq:TF}
        \mathcal{T}(k) = \Big\langle
        \frac{\mathcal{P}(\textbf{\textsf{M}}_{\rm clean}, \textbf{\textsf{M}})}{\mathcal{P}(\textbf{\textsf{M}}, \textbf{\textsf{M}})}
        \Big\rangle_{N_{\rm mock}} \, .
    \end{equation}
    \item Where $\mathcal{P}(\textbf{\textsf{X}}_{\rm a}, \textbf{\textsf{X}}_{\rm b})$ represents the operator that measures power between two fields (or a single field for an auto-correlation where a$\txteq{\equiv}$b) and bins into band powers (e.g.\ spherically averaged $k$-bins, although this is also applicable to cylindrical averaged binning). Then the measured power is corrected using 
    \begin{equation}\label{eq:Prec}
        P_{\rm rec}(k) = P_{\rm clean}(k)\,[\mathcal{T}(k)]^{-1} \, .
    \end{equation}
\end{enumerate}
This empirical approach accounts for the complex correlations and 
anti-correlations between signal and foreground modes, which would otherwise be neglected, leading to under-estimation of signal loss \cite[as demonstrated in][]{Cunnington:2023jpq}. The subtraction of the square-brackets term (i.e.\ the foreground-cleaned data) in \autoref{eq:Mclean} can be useful for reducing variance in the transfer function. However, further utility from the transfer function is to use it as a basis for covariance estimation (discussed in \secref{sec:SignalDetections}). In this case, it is important that the square bracket term not be subtracted so as not to under-estimate error bars.

\begin{figure}[!htb]
    \centering
    \includegraphics[width=0.465\linewidth]{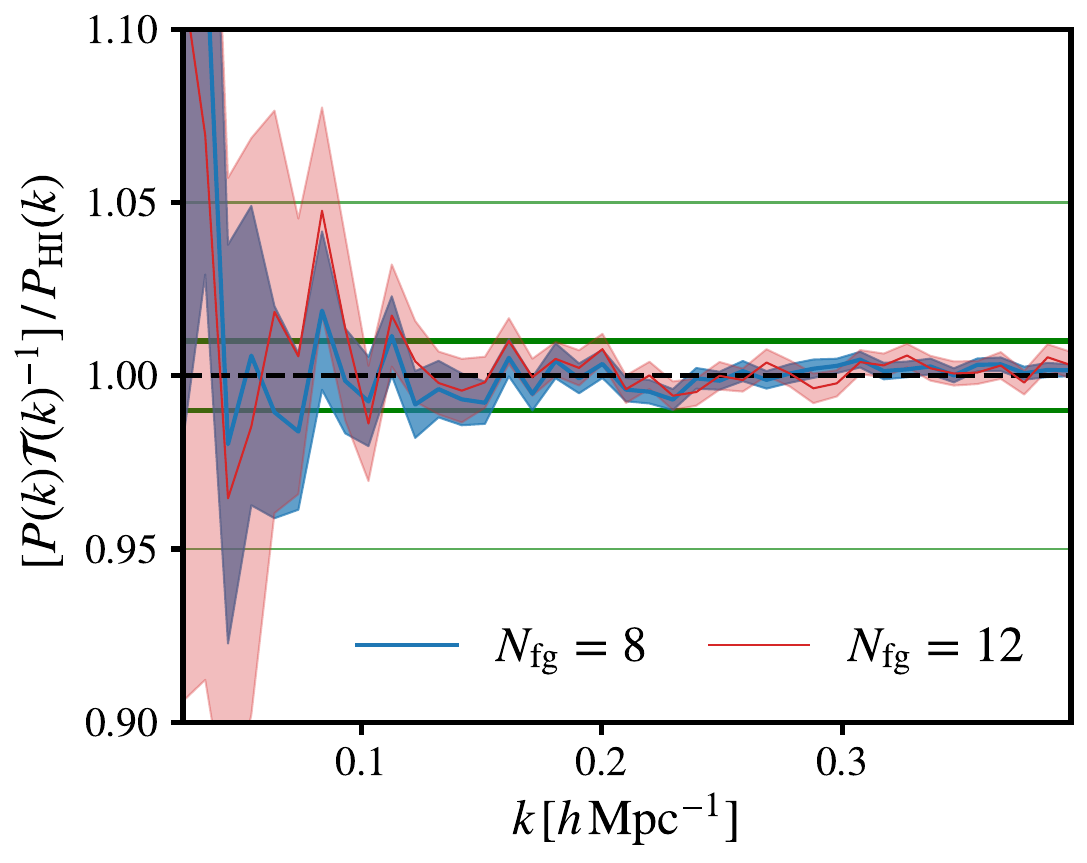}
    \includegraphics[width=0.525\linewidth]{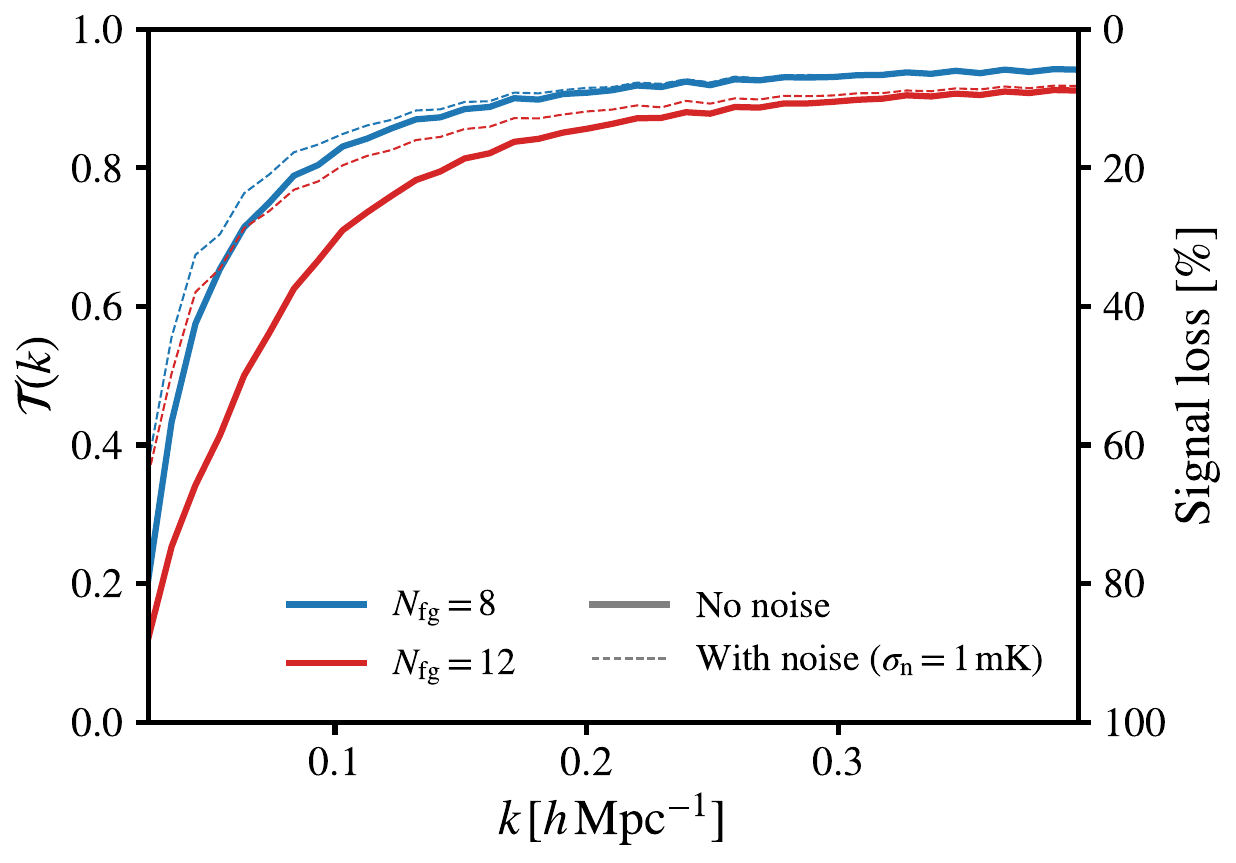}
    \caption{(\textit{left}): Accuracy of reconstructed foreground cleaned power spectra relative to the foreground-free \hi-only data ($P_\hi$) from simulated intensity maps. The transfer function, $\mathcal{T}(k)$, is constructed using \autoref{eq:TF} and 100 lognormal mocks, independent of the main $N$-body simulation assumed as the observational data. The shaded bands show the rms over these 100 mocks. A mild ($\Nfg\txteq{=}8$, blue line) and more aggressive ($\Nfg\txteq{=}12$, red line) foreground clean are shown. Dark-thick (light-thin) green horizontal lines indicate sub 1\% (5\%) accuracy regions of the reconstructed power spectrum. (\textit{right}): The shape of the transfer. Solid lines indicate noise-free simulations, thin dashed lines are where dominant white noise with rms $\sigma_\text{n}\txteq{=}1\,\text{mK}$ is added to the simulated observations.}
    \label{fig:TFdemos}
\end{figure}

Work from the MeerKLASS collaboration has shown that even aggressive cleaning (removing many modes) can be reliably corrected. The left panel of \autoref{fig:TFdemos} shows the reconstructed spherically averaged power spectrum is unbiased at the sub-percent level on intermediate and small scales, and within uncertainties on the largest scales. Here, a set of $N$-body semi-analytical simulations \cite[see][for full details]{Cunnington:2023jpq} was treated as the real ``observations'' with true power $P_\hi$. \textcolor{black}{Simulated foreground contamination was added, including diffuse Galactic synchrotron and free-free emission, bright point sources, along with some instrumental polarisation leakage to increase the spectral complexity. This foreground contaminated simulation was then cleaned using standard PCA, and then simpler lognormal \hi-mocks} were used to construct the transfer function $\mathcal{T}$. The known truth of the simulated data allows us to validate the ability of the transfer function to reconstruct the signal by comparing it to the original, foreground-free, power $P_\hi$. This was tested for two cases with $\Nfg\txteq{=}8$ and 12 PCA modes removed. Similar accuracy is achieved in both cases, with a slightly larger variance for $\Nfg\txteq{=}12$ due to the more aggressive cleaning.

The right panel of \autoref{fig:TFdemos} shows the shape of the transfer functions. The low $\mathcal{T}$ values at small-$k$ are expected and show that more signal is lost at large scales. \autoref{fig:TFdemos} also demonstrates the interesting impact of including purely Gaussian thermal noise in the simulation. In this high noise case ($\sigma_{\rm n}\txteq{=}1\,$mK), the noise dominates over the \hi\ fluctuations. The high noise causes a poorer foreground cleaning performance because foreground modes are more perturbed by the noise, thus more foreground contamination will remain in the residuals. This means signal loss is less drastic (comparing the dashed lines for the $\Nfg\txteq{=}8$ and 12 cases). This is particularly important for \textit{early} MeerKLASS, and other, intensity mapping surveys to consider where observational time may not have sufficiently accrued to suppress thermal noise. 

\medskip
\noindent
\textbf{Application in precision cosmology}:
A surprising and beneficial property of the transfer function is its weak dependence on the assumed cosmological model used in the injected mock simulations. In principle, the construction of the transfer function requires a fiducial cosmology to generate the input \hi\ power spectrum, and a mismatch between the assumed and true cosmology could in turn bias the recovered clustering statistics. However, it was empirically demonstrated in \citet{Cunnington:2023jpq} that such discrepancies yield only minimal errors in parameter inference. This robustness is particularly encouraging for ultra-large scale probes such as local primordial non-Gaussianity, $f_{\mathrm{NL}}$ \citep{Bartolo:2004if}, where the largest modes most affected by signal loss are also the most important for improved constraints.  

The origin of this robustness was clarified in \citet{Chen:2025til}, which showed using a quadratic estimator formalism that the transfer function is mathematically equivalent to a normalisation of the window function of the estimator, and therefore depends only on the linear operator that performs the foreground cleaning. In this picture, the transfer function can be interpreted as a renormalisation that also encapsulates mode-mixing induced by PCA cleaning, and is thus independent of the cosmology of the mock signal. The work in \citet{Chen:2025til} also provided a formal proof that, owing to the orthogonality of the PCA eigenmodes, the correction for \textit{both} cross- and auto-power is $\mathcal{T}^{-1}(\boldsymbol{k})$, contrary to the previously assumed $\mathcal{T}^{-2}$ for the auto-spectrum. 

\citet{Chen:2025til} also highlighted the important role mode-mixing (caused by the PCA) can have on the signal reconstruction. This revealed that the transfer function in its form defined by \autoref{eq:TF} and \autoref{eq:Mclean}, will include effects of mode-mixing, not just signal loss. To isolate the pure signal loss, and enforce further model independence, \textcolor{black}{an alternative approach}, where the eigenmodes are not recalculated with mock injection, can be taken, i.e.\ using the below in \autoref{eq:TF} instead:
\begin{equation}\label{eq:Mclean_var}
    \textbf{\textsf{M}}^\prime_{\rm clean} = \textbf{\textsf{M}}
        - \textbf{\textsf{V}}_{\rm obs}\, \textbf{\textsf{S}}\, \textbf{\textsf{V}}_{\rm obs}^{\top}\, \textbf{\textsf{M}}\, .
    \end{equation}
Without an additional correction for the mode mixing, this approach actually delivers a biased power spectra recovery \cite[shown in the upper panel of Figure C1 of][]{Cunnington:2023jpq}. However, as motivated by \citet{Chen:2025til}, it provides a more solid foundation to treat the complex effects of mode-mixing separately from signal loss. Hence, future MeerKLASS work will pursue this route for robust, model-independent signal loss correction, with increased suitability for precision cosmology.

\section{Regridding the sky into a Cartesian domain}\label{sec:Regridding}

A further challenge for intensity mapping analyses is the transformation of observational data from its native coordinates on the sky, $(\mathrm{R.A.}, \mathrm{Dec.}, \nu)$, into a three-dimensional Cartesian cube (x,\,y,\,z coordinates) in comoving units of $h^{-1}\mathrm{Mpc}$. This step is essential because estimators for clustering statistics are currently most efficiently applied in Fourier space, where Fast Fourier Transforms (FFTs) require data sampled on a regular grid \citep{Cooley1965AnAF}. Remaining in configuration space and using two-point correlation functions is a viable alternative \citep{Kennedy:2021srz,Avila:2021wih} with some useful properties, such as a more natural way to handle complex survey masks. However, as intensity mapping surveys scale up in size, counting pixel pairs will get increasingly computationally expensive. One could also, in principle, compute Fourier modes directly from sky voxels (without regular grid FFTs), but again the enormous size of modern surveys renders this computationally infeasible, making Cartesian regridding a critical prerequisite for Fourier-based (e.g.\ power spectra) analysis.  

Many recent pathfinder studies approximated the survey footprint as a rectangular box in comoving space, assigning comoving lengths to the map edges and then applying an FFT directly \citep{Wolz:2015lwa,COMAP:2021sqw}. However, this approximation introduces significant biases, with power spectrum estimates offset by more than $20\%$ across scales, and up to $30\%$ on the largest modes, even for surveys with relatively modest sky coverage. The root cause lies in the non-cubical geometry of real surveys: intensity mapping observations discretise the sky into voxels of unequal comoving volume, forming truncated conical footprints rather than rectangular boxes. Neglecting this geometry leads to systematic distortions in measured clustering, and this will worsen as surveys scale up in size.

\begin{figure}
    \centering
    \includegraphics[width=0.55\linewidth]{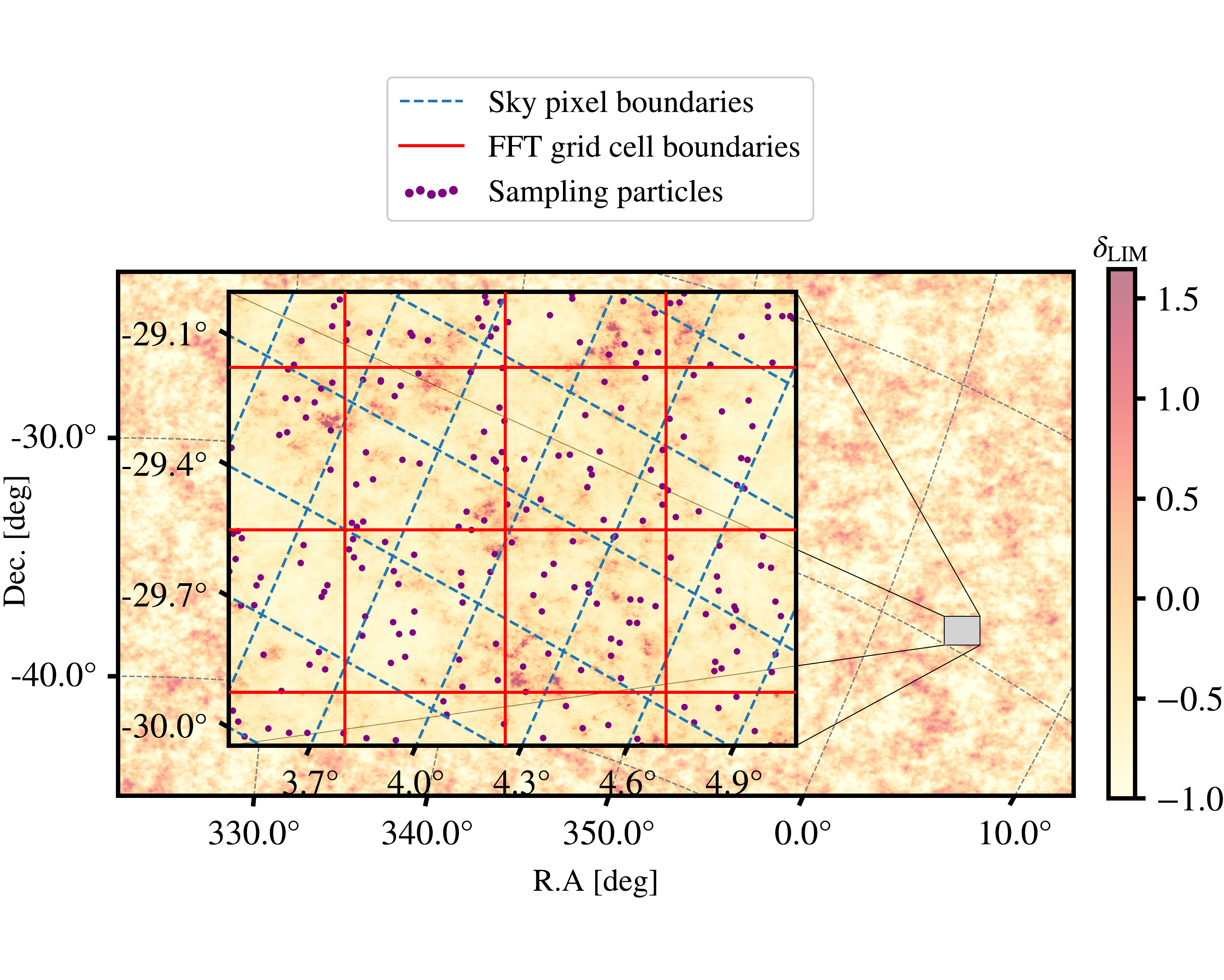}
    \includegraphics[width=0.44\linewidth]{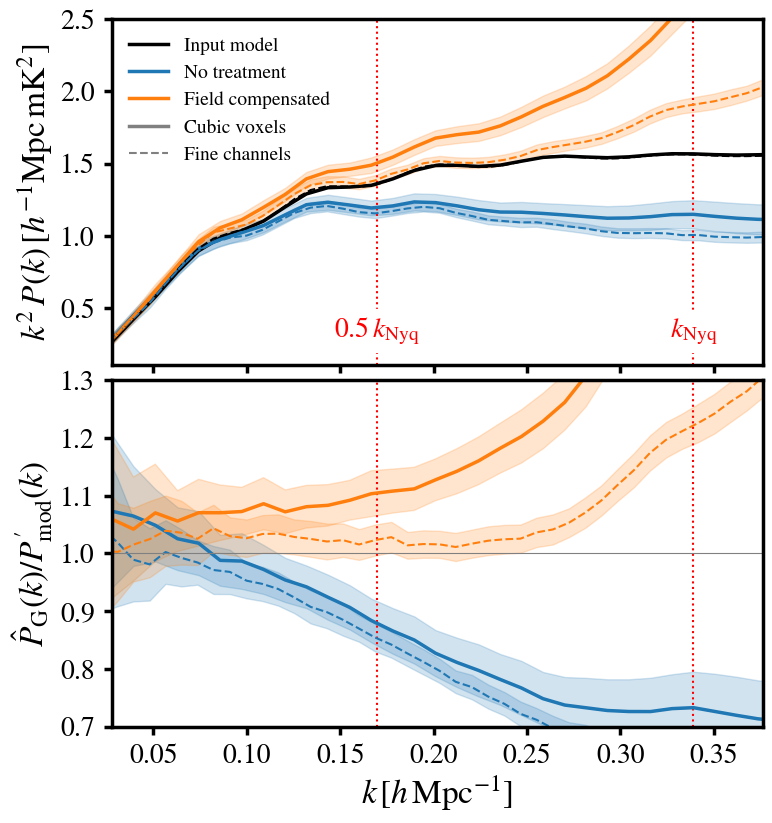}

    \caption{(\textit{left}): Pixelisation and Cartesian gridding processes necessary for intensity mapping analysis in Fourier-space. The background image represents a continuous fluctuation field. Intensity mapping surveys will discretise calibrated observations of the continuous field into broad sky pixels (blue dashed boundaries in insert). This provides the 3D voxel array in sky coordinates (R.A., Dec, $\nu$). Using random sampling particles, the sky voxel intensities can be regridded onto a 3D Cartesian grid (red boundaries) in comoving space ($l_\text{x}$, $l_\text{y}$, $l_\text{z}$ in $h^{-1}\text{Mpc}$). (\textit{right}): Accuracy of measured power spectrum from regridded fields relative to input model (black lines) for the two simulation versions (Cubic voxels and Fine channels). We use NGP interpolation for with no compensation (blue lines) and with compensation using \autoref{eq:Compensated} and \autoref{eq:W_ngp} (orange lines). The $1\sigma$ scatter from the 100 simulation realisations is shown by the shaded regions.}
    \label{fig:Gridding}
\end{figure}

To address this, MeerKLASS currently adopts the framework developed in \citet{Cunnington:2023aou}, with a particle resampling framework that redistributes the voxel intensities of the input sky maps into a uniform 3D Cartesian grid (see the left panel of \autoref{fig:Gridding} for a visual demonstration). By populating each sky voxel with sampling particles whose coordinates are transformed into comoving space, and then assigning these to regular grid cells, the method can avoid large-scale projection distortions.

Whilst this resampling will sufficiently sample the voxel intensities into a new \textcolor{black}{Cartesian} frame with minimal distortion, a statistical analysis of this regridded field alone will reveal significant biases relative to the true model. This is shown by the blue lines in the right panel of \autoref{fig:Gridding}. Here the estimated power spectra $\hat{P}_{\rm G}$ from the regridded intensity mapping simulations are compared to the corresponding input model $P_{\rm mod}$. Results are displayed for two different simulation versions, the ``Cubic voxels'' and ``Fine channel'', discussed more shortly. The power spectra are averaged over 100 simulation realisations \cite[see][for more details]{Cunnington:2023aou}. The vertical dashed red lines indicate the Nyquist frequency $k_{\rm Nyq}$ and $0.5\,k_{\rm Nyq}$ of the Cartesian grid.  

The blue curves show a clear suppression of power at high-$k$. This is caused by a smoothing to the field which is introduced by the nearest-grid-point (NGP) mass assignment of the resampling particles. Fortunately, this is a well-studied related problem in discrete galaxy catalogue-based clustering analyses \citep{Jing:2004fq,Sefusatti:2015aex}. To mitigate this, one can correct for the effective convolution of the field with the mass assignment window function, $W_{\rm mas}(\boldsymbol{x})$. This is most simply done in Fourier space by dividing the field through by the Fourier-space transform of the assignment function,
\begin{equation}\label{eq:Compensated}
    \tilde{\delta}_\text{G}(\boldsymbol{k}) \rightarrow \frac{\tilde{\delta}_\text{G}(\boldsymbol{k})}{\tilde{W}_\text{mas}(\boldsymbol{k})}\,.
\end{equation}
For NGP, effectively a top-hat in configuartion space, this corresponds to a product of sinc functions for each dimension in Fourier space \citep{HockneyEastwood}
\begin{equation}\label{eq:W_ngp}
    \tilde{W}_{\rm ngp}(\boldsymbol{k})= \text{sinc}\left(\frac{k_\text{x} H_\text{x}}{2}\right)
    \text{sinc}\left(\frac{k_\text{y} H_\text{y}}{2}\right)
    \text{sinc}\left(\frac{k_\text{z} H_\text{z}}{2}\right)\,,
\end{equation}
where $H\,{=}\,l/n_\text{G}$ is the cell width in the grid, which can be unique in each dimension. The orange curves in \autoref{fig:Gridding} show the result after applying this correction, revealing an enhancement of power towards the Nyquist scale. The over-estimation of power (particularly of the Cubic voxels simulation) is driven by aliasing, whereby unresolved small-scale fluctuations between particles fold into the accessible $k$-range and artificially boost the measured power spectrum.

The difference between the Fine channel simulations is that the spacing in frequency along the line-of-sight is much finer, more resembling a typical radio intensity mapping survey. In the Fine channel case (dashed lines), the increased frequency resolution suppresses aliasing in the radial modes, leading to smaller deviations from the true model compared with the Cubic voxels case (solid lines). \autoref{fig:Gridding} therefore demonstrates the two main systematic effects introduced by the regridding process: damping of small-scale power by particle assignment, and artificial enhancement by aliasing at high-$k$, the latter highlighting the need for further mitigation. This is also a problem shared by galaxy clustering analyses, and higher-order particle assignments and interlacing schemes exist \citep{Sefusatti:2015aex}, which can increase accuracy further if required. This was also explored in \citet{Cunnington:2023aou} and with targeted treatment, sub-percent accuracy could be achieved at up to 80\% of the Nyquist frequency.

In current analyses \cite[e.g.][]{MeerKLASS:2024ypg}, we do not see evidence of aliasing due to the relatively low signal-to-noise from these early data sets; therefore, simple NGP assignments with no interlacing are adopted for simplicity, but the smoothing correction to the regridded field from particle sampling (\autoref{eq:Compensated}) is implemented. However, for future MeerKLASS observations, which are beginning to cover large sky areas with high sensitivity, adopting robust regridding methods is crucial to avoid spurious power spectrum features and to unlock the full cosmological potential of the survey. 

\subsection{Analysis beyond the flat-sky approximation}

The current approach of regridding does not fundamentally escape the problem of a flat-sky approximation. In regridded fields, we must assume a global line-of-sight direction, and for a wide-field survey, this will not be mutually parallel with all lines-of-sight. To date, even the most advanced galaxy surveys \cite[see e.g.][]{DESI:2025zgx} have relied on a flat-sky (or plane-parallel wave) approximation, embedding survey footprints into a Cartesian cube and applying ad-hoc wide-angle corrections. While this approach is workable for modest survey areas, its accuracy breaks down as coverage extends to wider areas, with biases creeping in for parameter inference (e.g.\ \fNL) on the largest scales \citep{Benabou:2024tmn,2025JCAP...07..063G}. For intensity mapping, the shortcomings are magnified because of the strongly anisotropic observational effects: frequency-dependent telescope beams that convolve purely \textit{angular} modes, and foreground cleaning that preferentially removes large-scale \textit{radial} modes. Accurately modelling or correcting these within a plane-parallel box will demand complex forward modelling and risks residual biases that undermine precision cosmology. A wide-angle correction recipe that also includes beam and similar configuration-space convolution effects is not currently available, but represents a viable option, whose effectiveness is currently being investigated within the MeerKLASS Collaboration. 

One alternative that circumvents many of these issues is to analyse the data in harmonic space, measuring angular power spectra $C_\ell$ as a function of frequency or redshift \citep{Asorey:2012rd,Camera:2018jys}. This formalism naturally respects the curvature of the sky, avoids plane-parallel assumptions, and has been widely used with great success \citep{Alonso:2018jzx,Tanidis:2020byi,Tanidis:2021uxp,Nakoneczny:2023nlt,DES:2024oud,Zheng:2025lqz}. Applied to intensity mapping, it would provide a clean framework to handle beam convolution and sky masks. However, harmonic analyses are intrinsically limited to angular clustering, effectively projecting out radial information within each frequency slice, unless a large number of radial bins are used, which then pushes the limits of computational feasibility \citep{Camera:2018jys,Semenzato:2024rlc}. As a result, the exquisite tomographic potential of intensity mapping, its ability to probe many frequencies across a wide redshift range, is not fully exploited. Similarly, line-of-sight-only $P(\nu)$ power spectra are feasible \cite[e.g.][]{Villaescusa-Navarro:2016kbz} and avoid regridding and plane-parallel approximations, but will face similar limitations, and will not fully exploit the full volume. This loss of three-dimensional information in $C_\ell$ and $P(\nu)$ approaches is particularly restrictive for ultra-large-scale cosmology, such as tests of primordial non-Gaussianity, which are optimal when they access the full radial and angular mode spectrum over vast cosmic volumes. 

Another intriguing solution is to use a fully 3D spherical treatment of the data. This is possible with the spherical Fourier–Bessel formalism \citep{Liu:2016xzv,Castorina:2017inr}, which provides a native decomposition into angular and radial modes, automatically accommodating wide-angle terms and respecting the true geometry of the observations. Importantly, it also allows observational operators, such as the chromatic beam or foreground cleaning filters, to be applied where they physically act, either on angular or radial modes, avoiding the distortions introduced by Cartesian embedding. Early theoretical work has already highlighted the potential of SFB methods to recover large-scale clustering unbiasedly \citep{Wen:2025zbv}, plus there is potential for improving the computational feasibility of this approach \citep{Wang:2020wsx,Raccanelli:2023zkj}. However, a robust end-to-end pipeline tailored to 21cm intensity mapping has yet to be developed. This is far beyond the scope of this work, but it is a planned exploration within the MeerKLASS collaboration.

\section{Signal detections}\label{sec:SignalDetections}

The 21cm line of \hi\ offers a uniquely powerful probe of cosmology across an extensive redshift range, from the relatively recent Universe deep into the epoch of reionisation (EoR) \citep{Zaroubi:2012in,Kovetz:2017agg}. Much of the focus from the 21cm intensity mapping community is dedicated to this high-redshift frontier, aiming to detect the imprint of reionisation on the distribution of \hi. However, such measurements are extremely challenging: foregrounds are even more overwhelming at these low frequencies \textcolor{black}{(hundreds of times brighter at $\nu\txteq{<}200\,$MHz compared to $1{,}000{\,}$MHz}), and despite progressive improvements on upper limits using instruments such as LOFAR, MWA, and HERA \citep{Mertens:2025pvk,Nunhokee:2025jbn,HERA:2022wmy}, a full detection is yet to be obtained.

By contrast, one of the major strengths of 21cm intensity mapping at lower redshifts, in the post-EoR Universe, is the ability to probe the large-scale distribution of matter in synergy with other cosmological tracers. In particular, cross-correlation with optical galaxy surveys provides a robust pathway to detection: foregrounds and systematics in the radio maps are uncorrelated with optical data, while the cosmological signal is shared between the two \citep{Wolz:2015ckn,Berti:2023viz,Santos:2025fff}. This makes cross-correlation a powerful validation tool as well as a means to extract the first cosmological information from current-generation surveys, and characterise additive systematics.  

In this section, we review some of the first detections of cosmological clustering in cross-correlation between early MeerKLASS intensity maps and overlapping galaxy surveys. These results represent a major milestone, demonstrating the feasibility of \hi\ intensity mapping for cosmology and paving the way for future measurements with the full MeerKLASS and, ultimately, the SKAO \citep{SKA:2018ckk}.

\subsection{Cross-correlation power spectra with optical galaxies}

The first detection of \hi\ intensity mapping with MeerKAT was reported in \citet{Cunnington:2022uzo} using cross-correlation with galaxies from the WiggleZ Dark Energy Survey. With just 10.5 hours of L-band single-dish observations from the MeerKAT pilot survey covering ${\sim}\,200\,$deg$^2$ in the WiggleZ 11hr field, a $7.7\sigma$ detection was obtained of the cross-correlation power spectrum. This provided a measurement of the combined parameter $\Omega_\hi b_\hi r\txteq{=}[0.86 \txteq{\pm} 0.10 \,{\rm (stat)} \txteq{\pm} 0.12 \,{\rm (sys)}] \txteq{\times} 10^{-3}$ at an effective scale of $k_{\rm eff}\txteq{\sim}0.13\,h\,{\rm Mpc}^{-1}$. Crucially, this result represented the first practical demonstration of the \textcolor{black}{the single-dish mode employed on a multi-dish instrument for cosmological intensity mapping}, establishing MeerKAT as a pathfinder for SKAO-era science.  

Building on this, the MeerKLASS collaboration reported results from the L-band deep-field observations in \citet{MeerKLASS:2024ypg}. These maps, produced from 41 repeated scans over 236\,deg$^2$ (totalling 62 hours per dish before flagging), yielded the deepest (in terms of time integration) single-dish \hi\ intensity maps to date. Cross-correlating with a smaller overlapping spectroscopic sample from the GAMA survey, only covering about one third of the field, a detection of the cross-power spectrum was obtained at ${>}\,4\sigma$ significance in the redshift range $0.39\txteq{<}z\txteq{<}0.46$. This work showcased substantial improvements in calibration, building upon the work in \citet{Wang:2020lkn} by including an iterative self-calibration. This work oversaw the transition into a regime where the thermal noise is no longer dominant compared to cosmological \hi\ fluctuations, necessitating more sophisticated covariance estimation strategies, which we discuss in the next sub-section.

Most recently the work in \citet{Carucci:2024qpm}, revisited the pilot survey data overlapping with WiggleZ but focused on advancing the contaminant separation strategy. Introducing and testing a novel multiscale foreground-cleaning approach (mPCA, introduced in \secref{sec:PCA}), this work demonstrated how a more efficient foreground cleaning can be more robust to signal loss. Without signal loss \textcolor{black}{reconstruction}, this pipeline was able to confirm the cross-correlation detection at ${\sim}\,6\sigma$ confidence, measuring $\Omega_\hi b_\hi r \txteq{=} [0.93 \txteq{\pm} 0.17] \txteq{\times} 10^{-3}$ robustly across a wide range of scales ($0.04 \txteq{\lesssim} k \txteq{\lesssim} 0.3\,h\,{\rm Mpc}^{-1}$). This result instils greater confidence that MeerKAT \hi\ intensity mapping analyses can deliver unbiased cosmological information, laying further groundwork towards the full MeerKLASS.

\begin{figure}
    \centering
    \includegraphics[width=0.9\linewidth]{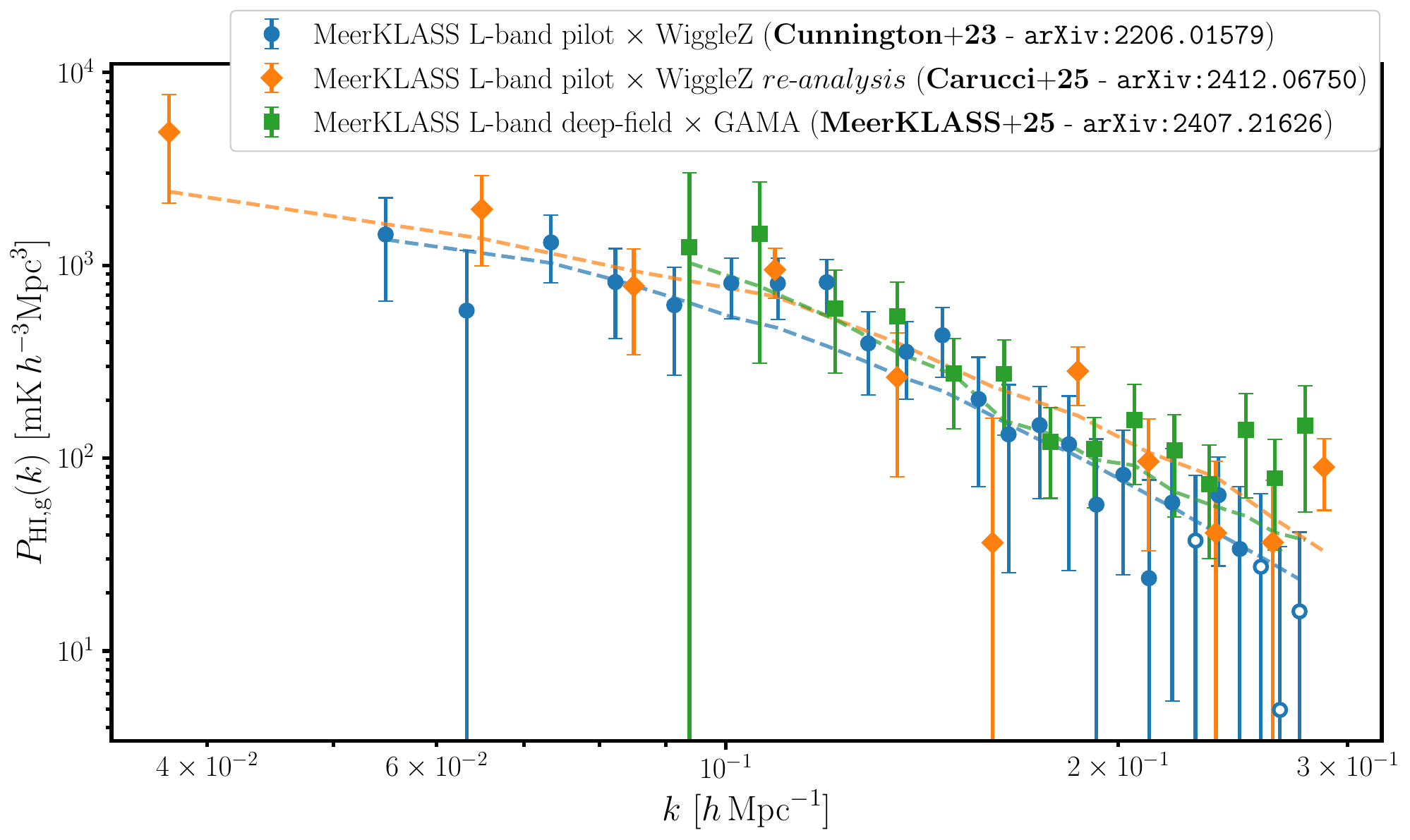}
    \caption{Cross-correlation power spectra between MeerKAT L-band intensity maps and optical galaxy surveys at $z\txteq{\sim}0.43$. \textit{Blue-circles} show the initial pilot survey detection with WiggleZ galaxies \citep{Cunnington:2022uzo} (hollow markers indicate negative power), \textit{orange-diamonds} the re-analysis of this data with improved foreground cleaning \citep{Carucci:2024qpm}, and the \textit{green squares} is the deep-field cross-correlation with GAMA \citep{MeerKLASS:2024ypg}. Error bars show $1\sigma$ uncertainties estimated using different methods in each detection. The dashed lines are the corresponding model matched by colour. We refer the interested reader to the dedicated papers for details on these subtle distinctions.}
    \label{fig:crossPks}
\end{figure}

\autoref{fig:crossPks} shows the three published cross-correlation power spectra overlaid on the same axes. The results are in good agreement within the signal-to-noise of the individual measurements, providing a consistent picture of the underlying \hi\ clustering. The \textcolor{black}{smaller range of scales} probed in the GAMA cross-correlation \citep{MeerKLASS:2024ypg} is a consequence of the much smaller survey area of GAMA relative to WiggleZ, which limits the accessible Fourier modes. The extended reach to smaller $k$ achieved by \citet{Carucci:2024qpm} arises from the more controlled multiscale foreground cleaning, which better preserves large-scale modes. All of these measurements were performed in the MeerKAT L-band, where RFI confines the usable redshift range to $0.39\txteq{\lesssim} z \txteq{\lesssim} 0.46$. Future observations in the UHF band (outlined in \secref{sec:Observations}) will dramatically expand the accessible redshift volume, allowing for much higher signal-to-noise detections and substantially improved cosmological constraints, which we discuss in more detail later in \secref{sec:Future}.  

\subsection{Covariance estimation and scale cuts} 

\begin{figure}
    \centering
    \includegraphics[width=1\linewidth]{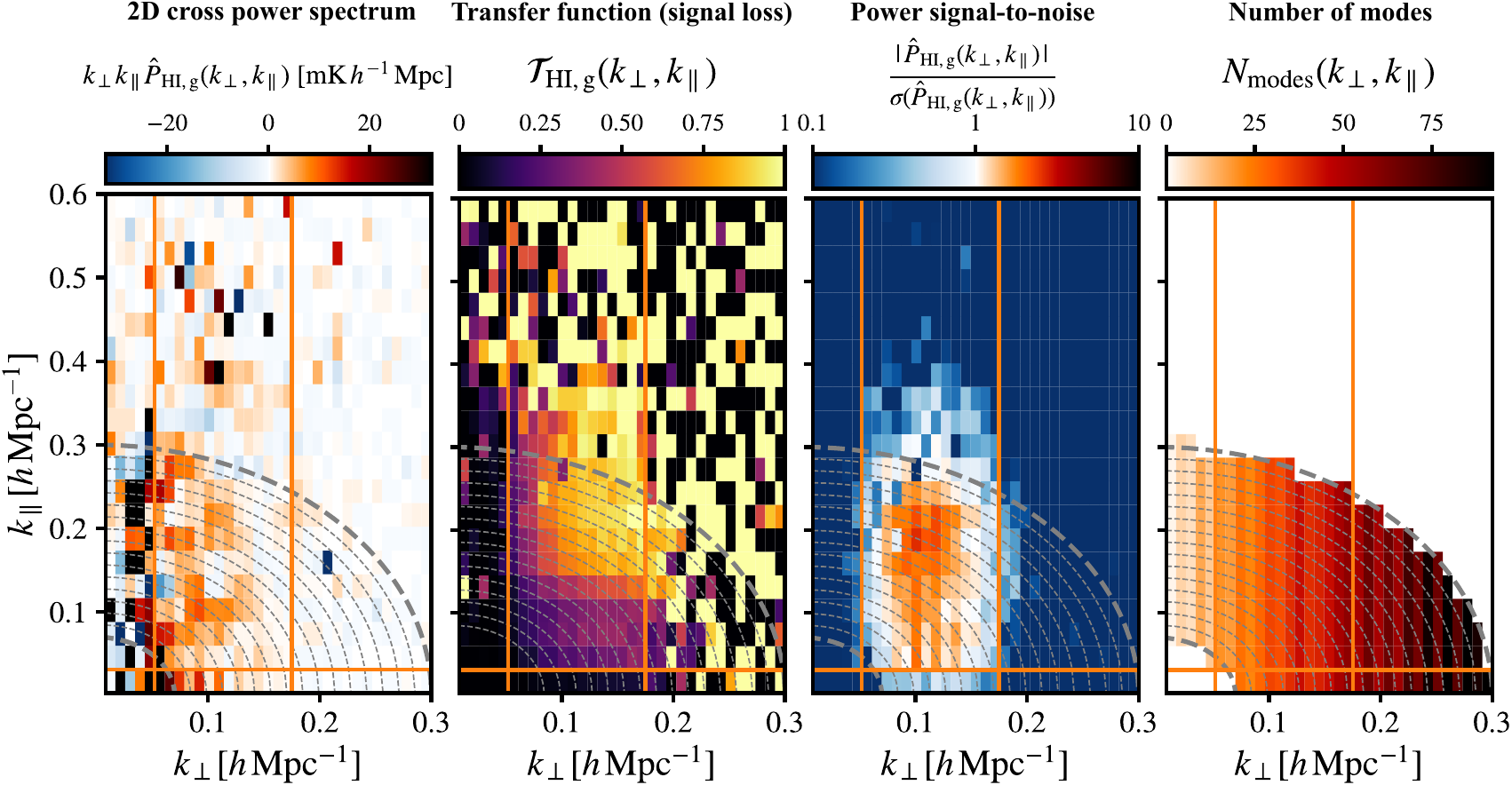}
    \caption{Two-dimensional cross-power analysis between the MeerKLASS L-band deep-field intensity maps and GAMA galaxies. The panels show (from left to right) the measured cross-power (with signal loss correction), the computed foreground transfer function, the estimated signal-to-noise, and the number of modes contributing to each $(k_\perp, k_\parallel)$ bin. The orange lines indicate the chosen $k$-cuts used to exclude poorly constrained regions of Fourier space from the spherically averaged $P(k)$ measurement (green results in \autoref{fig:crossPks}), and the grey contours show the $k$-bin boundaries chosen for this $P(k)$.}
    \label{fig:2DTF_SNR}
\end{figure}

The work in \citep{MeerKLASS:2024ypg} investigated in detail how to efficiently extract the cosmological signal from MeerKLASS data and also estimate the errors on any measurements. The far-left panel of \autoref{fig:2DTF_SNR} again shows the cross-correlation power spectrum between the MeerKLASS deep field and the GAMA galaxies but, instead of the 1D spherically-averaged power, this is \textit{cylindrically} averaged into ($k_\perp$, $k_\parallel$) bins. This gives a \textcolor{black}{more detailed view} into where the signal exists in this analysis. The panel clearly shows the strong effect of the instrumental beam, where the power on small transverse scales (large $k_\perp$) is suppressed to nearly zero. At the opposite end, the strength of the signal at extremely small $k$ is also unreliable, which can be attributed both to survey geometry limitations and to residual contamination from foregrounds. This interpretation is supported by the transfer function analysis (shown in the second panel), which indicates that nearly all of the cosmological signal is lost ($\mathcal{T}\txteq{\sim}0$) at the very smallest $k$-modes owing to their strong degeneracy with foreground spatial structure. 

A major part of the investigation in \citep{MeerKLASS:2024ypg} was to explore the use of the transfer function not only as a correction for signal loss but also as a means of estimating the covariance of the power spectrum. In earlier analyses \citep[e.g.][]{Cunnington:2022uzo}, analytical error estimates were adopted, assuming that thermal noise dominated the error budget and that Gaussian statistics applied. However, with the improved sensitivity of MeerKLASS, these assumptions no longer hold: thermal noise is subdominant on many scales, residual systematics are present, and variance from the PCA foreground cleaning is not accounted for.  

An alternative approach is to use the scatter of the transfer function across the ensemble of mock injections as an error estimator. The cross-correlation transfer function for a single $i$th mock is defined as
\begin{equation}
    \mathcal{T}_i(k) = \frac{\mathcal{P}(\textbf{\textsf{M}}_{\rm clean}, \textbf{\textsf{M}}_\text{g})}{\mathcal{P}(\textbf{\textsf{M}}_\hi, \textbf{\textsf{M}}_\text{g})} \, .
\end{equation}
where $\textbf{\textsf{M}}_{\rm clean}$ is as before in \autoref{eq:Mclean}, but here we do \textit{not} subtract the $\textbf{\textsf{X}}_{\rm clean}$ square bracket term, since this includes a lot of the variance in the power spectrum measurement we are now aiming to quantify. To isolate the contribution that drives the error, we define the transfer function fluctuation for each mock by subtracting the ensemble mean over all mocks
\begin{equation}
    \delta \mathcal{T}_i(k) = \mathcal{T}_i(k) - \langle\mathcal{T}(k)\rangle_{N_{\rm mock}} \, .
\end{equation}
Assuming independent (quadrature) addition of uncertainties for the \autoref{eq:Prec} ratio $\hat P_{\rm rec}=\hat P_{\rm clean}/\mathcal{T}$, standard error propagation suggests
\begin{equation}
    \frac{\delta \hat P^{\rm rec}_i(k)}{\hat P^{\rm rec}(k)} = 
    \sqrt{
    \left(\frac{\delta \hat P^{\rm clean}_i(k)}{\hat P^{\rm clean}(k)}\right)^2
    +
    \left(\frac{\delta \mathcal{T}_i(k)}{\mathcal{T}(k)}\right)^2
    } \, .
\end{equation}
In this transfer function scatter approach, we assign all quantifiable uncertainty to the transfer function distribution itself (the measured $\hat P_{\rm clean}$ is held fixed across the ensemble), so $\delta \hat P^{\rm clean}_i\txteq{=}0$. This yields the working relation
\begin{equation}
    \delta \hat P^{\rm rec}_i(k)
    \;=\;
    \hat P^{\rm rec}(k)\,
    \frac{\delta \mathcal{T}_i(k)}{\mathcal{T}(k)} \, ,
\end{equation}
i.e.\ the fluctuation of the reconstructed power for each mock is directly proportional to the transfer function fluctuation for that mock. 

This provides a distribution of $\delta \hat{P}_i^{\rm rec}$ uncertainties on which 68th percentile limits can be taken to provide $1\sigma$ error bar estimates. Furthermore, the covariance matrix can be trivially estimated from this distribution by taking the covariance over $\hat P^{\rm rec}_i(k) \txteq{=} \hat P^{\rm rec}(k) \txteq{+} \delta \hat P^{\rm rec}_i(k)$. These estimates will naturally include contributions from sample variance, galaxy shot noise (from the galaxy mocks which replicate the real galaxy density), residual foregrounds and systematics, instrumental noise, and the variance associated with signal loss from eigenmode projection. Importantly, because the mocks are injected into the real data, any residual contaminants that persist in the cleaned maps are automatically included in the covariance budget \textcolor{black}{(which we discuss in more detail shortly)}. 

\textcolor{black}{The covariance and error estimations} can be computed in the spherical- or cylindrical-averaged power spectrum space. This forms the basis for estimating the signal-to-noise of the cross-power as shown by the third panel of \autoref{fig:2DTF_SNR}. This shows the corresponding signal-to-noise distribution, highlighting which regions of $(k_\perp, k_\parallel)$ space are robustly measured. This directly motivated the adoption of scale cuts, illustrated by the orange lines, where modes outside the well-behaved region are excluded from the final spherically-averaged power spectrum. As the fourth panel demonstrates, these cuts remove a large chunk of the available Fourier modes. Nevertheless, this conservative strategy is necessary with the current survey volume and sensitivity. In the future, as MeerKLASS expands in area and as larger samples become available with the SKAO, many more modes will be accessible, and the statistical penalty from scale cuts will be less severe.

\begin{figure}
    \centering
    \includegraphics[width=0.49\linewidth]{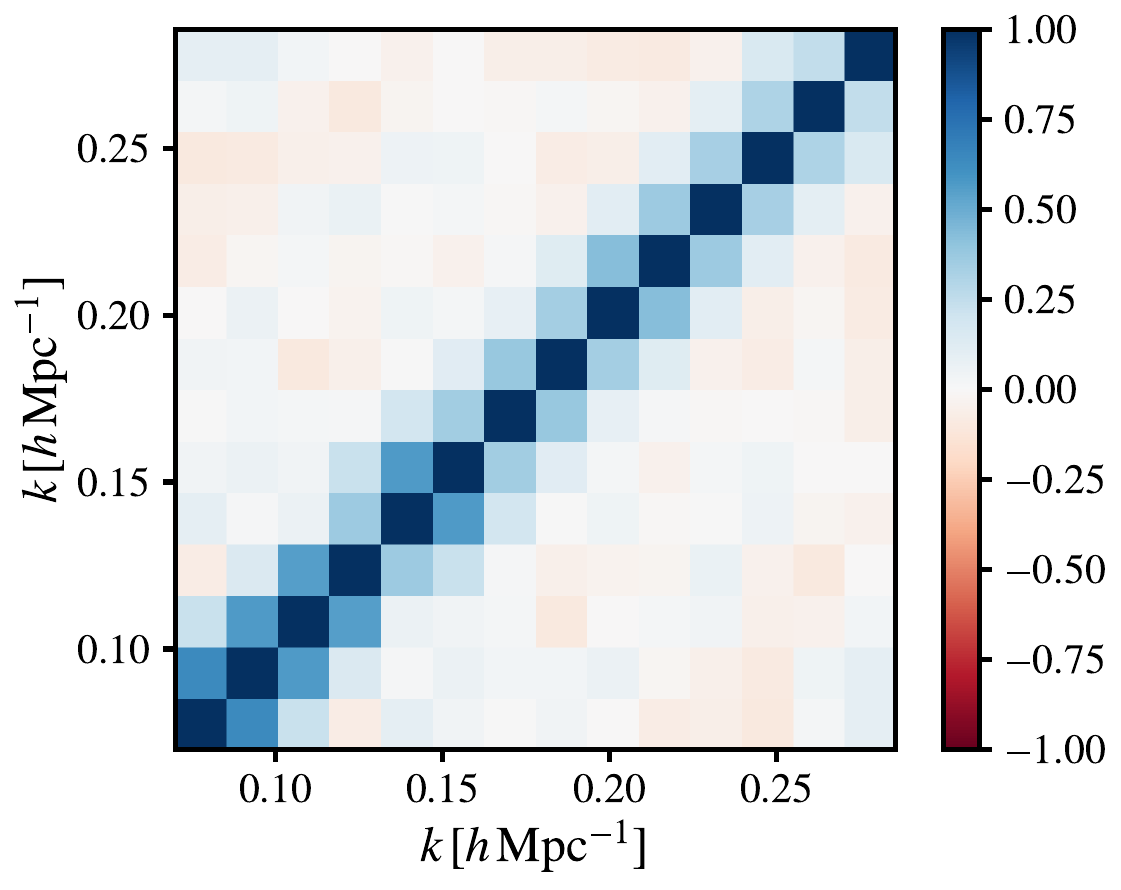}
    \includegraphics[width=0.49\linewidth]{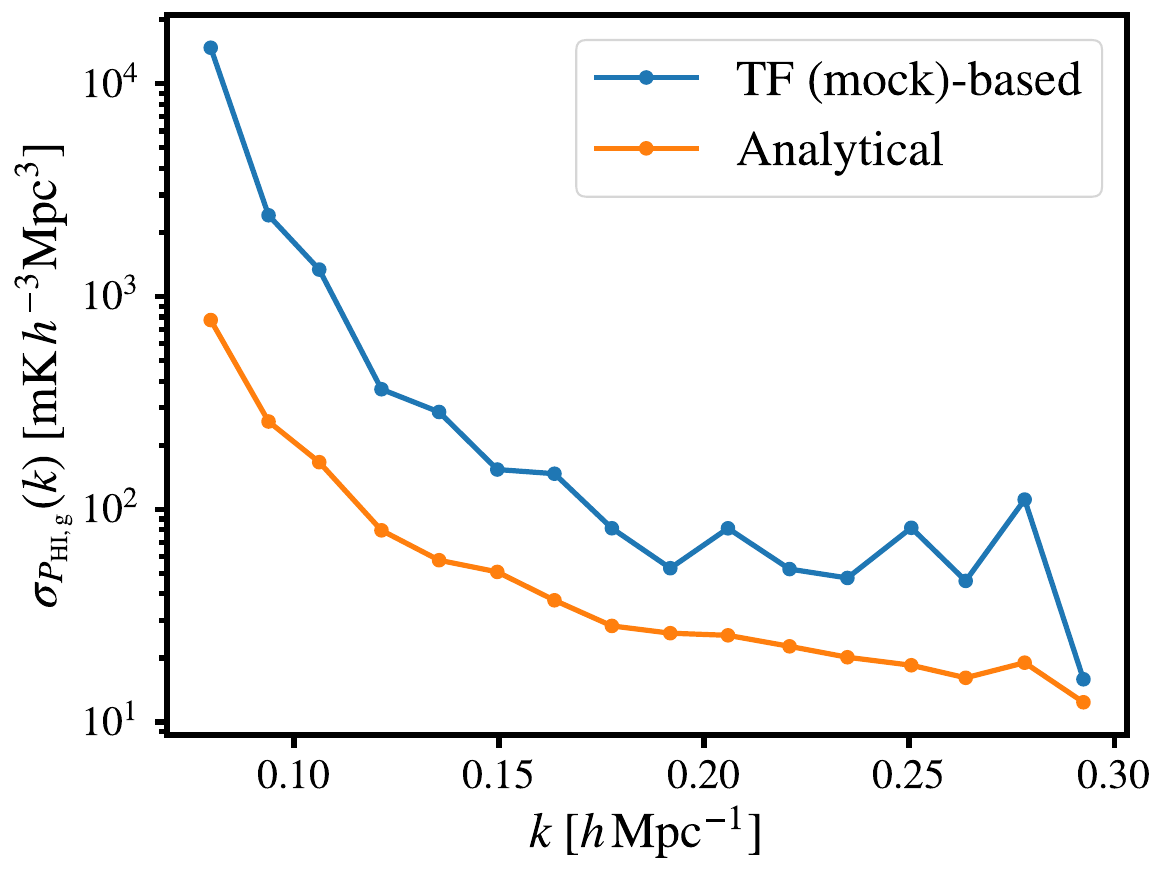}
    \caption{\textcolor{black}{(\textit{left}): $k$-bin correlation matrix for the cross-power between the MeerKLASS L-band deep-field intensity maps and GAMA galaxies, following the semi-empirical approach that utilises the transfer function variance (detailed in the text). (\textit{right}): Diagonal covariance comparison between the transfer-function based approach (blue) computed from the diagonals of the full covariance matrix, and an analytical estimation (orange), outlined in the text, used in previous analyses.}}
    \label{fig:covariance_kbin}
\end{figure}

\medskip
\noindent
\textbf{More robust error estimation, more conservative results}: \textcolor{black}{The normalised covariance matrix, i.e. the correlation matrix, is shown in the left panel of \autoref{fig:covariance_kbin} for the MeerKLASS deep-field cross power measurement. This uses the empirical transfer-function-based approach, and we clearly see non-negligible contributions from the off-diagonal elements, motivating this extension beyond the purely analytical diagonal approach in previous analyses.}

An outcome of the analysis in \citet{MeerKLASS:2024ypg} was that the estimated errors on the measured cross-power spectrum were larger than anticipated, and certainly larger than analytical methods used in the earlier intensity mapping detections \citep{eBOSS:2021ebm,Cunnington:2022uzo}. \textcolor{black}{We demonstrate this in the right panel of \autoref{fig:covariance_kbin} where we compare the (\textit{blue}) diagonals of the transfer-function-based covariance (i.e. a mock-based calculation with empirical contributions from real systematics in the data) and (\textit{orange}) an analytical estimation of the diagonal (Gaussian) covariance calculated from}
\begin{equation}\label{eq:theory_error}
    \textcolor{black}{\sigma_{P_{\hi, \mathrm{g}}}(k)=\frac{1}{\sqrt{2 N_{\mathrm{modes}}(k)}} \sqrt{\hat{P}_{\hi, \mathrm{g}}^2(k)+\hat{P}_{\hi}(k)\hat{P}_{\mathrm{g}}(k)}\,.}
\end{equation}
\textcolor{black}{Here, $\hat{P}_\hi$ and $\hat{P}_\text{g}$ are the estimators for the \hi\ and galaxy auto-correlation power spectra, inclusive of the thermal noise and galaxy shot-noise, respectively.}

The mock-based transfer function method naturally captures contributions from residual systematics, the variance in signal loss due to eigenmode projection, and the presence of correlated errors between bins in Fourier space. This contrasts with earlier diagonal covariance assumptions, where only independent errors were derived based on auto-correlation noise estimation \textcolor{black}{(see \autoref{eq:theory_error})}. Including these additional contributions leads to a more realistic, but inevitably more conservative, characterisation of the significance of the signal. Furthermore, the work in \citet{Chen:2025til}, which, as discussed, revealed that the transfer function is currently complicated by mode-mixing, could also be biasing the covariance estimation. Hence, further work is planned (as discussed in \secref{sec:TF}) to investigate the numerical reconstruction routines and settle on the most robust approach for covariance estimation.

As a result, the detection of the GAMA cross-correlation is quoted at the ${\sim}\,4\sigma$ level, compared to the ${\sim}\,7\sigma$ significance claimed for the WiggleZ field. The larger field size of WiggleZ also makes this a less direct comparison. However, it is important to stress that this more conservative result does not necessarily imply a weaker underlying signal. Instead, it reflects the fact that the new pipeline is incorporating variance at a more complete level, albeit possibly being overestimated due to mode-mixing \citep{Chen:2025til}. Indeed, it was shown that if the contribution from systematics could be eliminated, driving down the covariance to the level expected for a Gaussian signal+noise (i.e.\ a $\chi^2$ consistent with uncorrelated bins and error bars representative of scatter), the GAMA-cross detection significance would increase to nearly $10\sigma$. This highlights both the robustness of the current conservative detection and the major potential gains that will be unlocked by continued improvements in calibration and systematic control.

\subsection{Pursuit of a HI auto-correlation power spectrum detection}

Previous single-dish intensity mapping experiments have been dominated by thermal noise at all scales, but with MeerKLASS we are now crossing the threshold where the cosmological \hi\ fluctuations begin to dominate the expected thermal noise on large scales with the L-band deep field observations.

\begin{figure}
    \centering
    \includegraphics[width=0.5\linewidth]{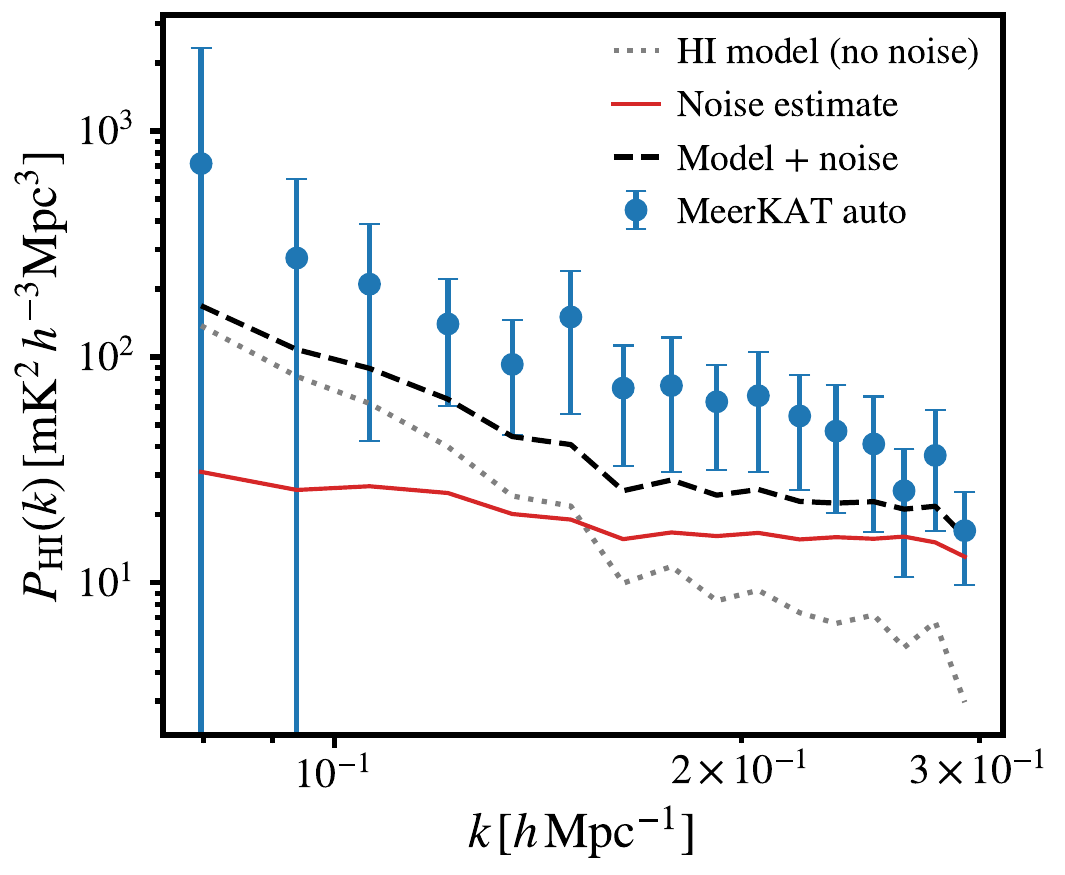}
    \caption{The \hi\ auto-power spectrum for the MeerKLASS L-band deep-field from \citep{MeerKLASS:2024ypg}. The red solid line is the noise estimate from measuring the average power of 10 Gaussian random fields with an RMS expected to be consistent with the thermal noie level. The grey dotted line is the predicted \hi\ power spectrum model and the black dashed line is the combination of this \hi\ model plus the thermal noise estimate.}
    \label{fig:HIauto_power}
\end{figure}

\autoref{fig:HIauto_power} shows \textcolor{black}{the MeerKLASS \textit{auto}-power spectrum from the deep field (blue data points) compared to a modelled prediction of a \hi-only auto-power (grey dotted line), and an estimate of the thermal noise (red solid line) obtained from Gaussian random realisations using the known observational time within each map pixel. On large scales ($k\txteq{\lesssim}0.15\,h{\rm Mpc}^{-1}$). The cosmological \hi\ power is predicted to exceed the thermal noise, and the total expected power (black dashed line) can be approximated by the sum of the two contributions. The data follow this expectation closely, with a small (${\sim}\,2$) mean residual factor between the model and the MeerKLASS measurements across all $k$.} This represents the tightest upper limit to date on the \hi\ auto-power on large scales (a detection of a \hi\ auto power spectrum has been reported in \citet{Paul:2023yrr} on smaller scales, \textcolor{black}{and now also in \citet{CHIME:2025cee}, again on smaller scales}).

The MeerKLASS collaboration is pursuing several routes for improvement in future work, including splitting the data into independent sub-sets (different dishes and scan times) to isolate and mitigate systematics. A more dedicated auto-correlation study will be presented in upcoming work \citep{MatildeAutoPk}, which will build upon this first result and push towards a robust measurement of the \hi\ auto-power.

\subsection{Stacking HI emission onto galaxy positions}

Another complementary route to extracting \hi\ information from MeerKLASS intensity maps is through spectral line stacking. In this technique, galaxy positions from an external redshift survey are used as priors, and sub-cubes of the 21\,cm map centred on these positions are averaged together. While the \hi\ emission from any individual galaxy lies far below the noise limit of current single-dish maps, stacking over thousands of positions builds up sufficient signal-to-noise to probe the average emission properties of the population.

\begin{figure}
    \centering
    \includegraphics[width=0.5\linewidth]{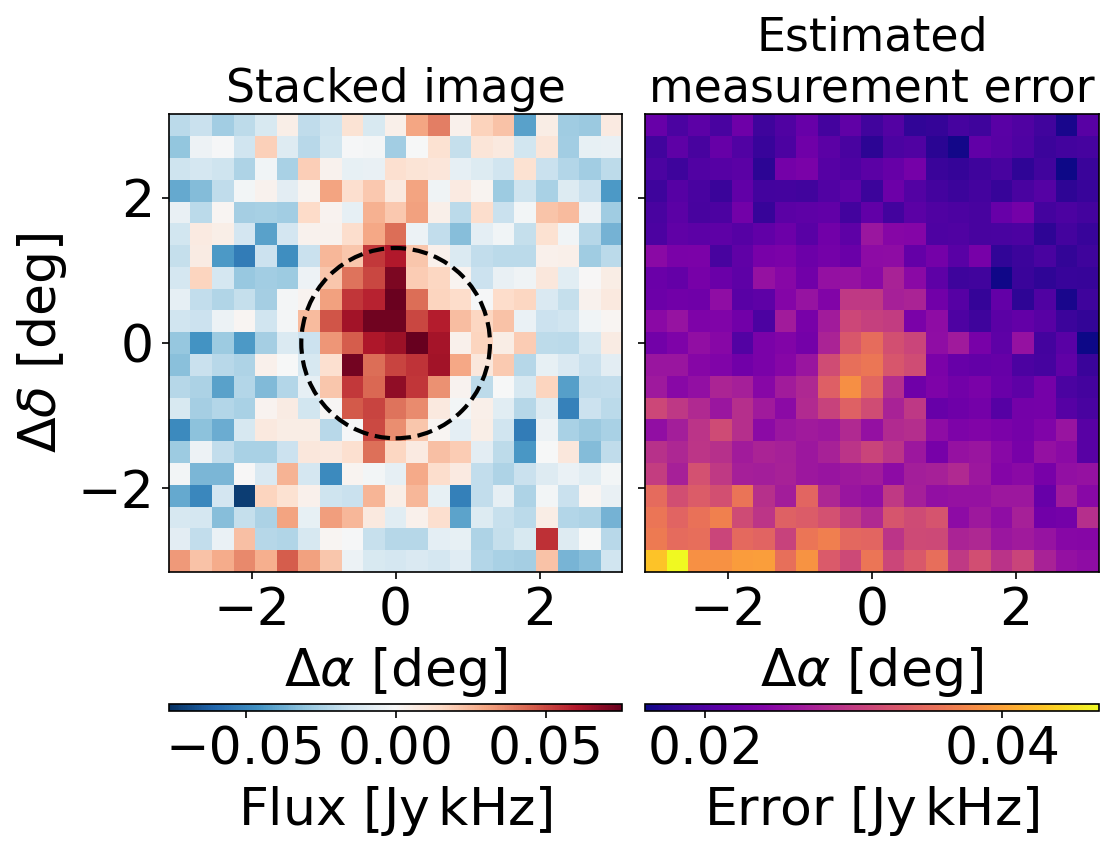}
    \caption{The emission line stacking measurement of MeerKLASS \hi\ intensity maps onto positions of GAMA spectroscopic galaxies. The left panel shows the detection of the emission line signal. The dashed line denotes the primary beam field-of-view. The right panel shows the estimated $1\sigma$ measurement error. }
    \label{fig:stack}
\end{figure}

The first demonstration of this approach with MeerKLASS was presented in \citet{MeerKLASS:2024ypg} with a more detailed follow-up and extension in \citet{Chen:2025bcx}. These studies showed that by stacking the L-band deep-field maps at the positions of GAMA galaxies, a clear ($7.45\sigma$) excess signal in the stacked profile was revealed, as shown in \autoref{fig:stack}. However, the work also highlighted challenges that are specific to broad-beam single-dish intensity mapping: the ${\sim}\,1^\circ$ beam implies that multiple galaxies fall within each resolution element, leading to double-counting and a plateau of excess emission in the stacked spectrum. Furthermore, the PCA cleaning applied to remove foregrounds introduces signal loss and spectral distortions that must be accounted for with forward modelling. Despite these complications, these studies provided the first evidence that stacking can recover meaningful information from single-dish intensity maps, complementary to power spectrum analyses.

The follow-up analysis in \citet{Chen:2025bcx} significantly extended this framework. Using the same MeerKLASS L-band deep-field data, the work presented a forward-modelling pipeline that incorporates clustering beyond Poisson statistics, beam convolution, and the effects of foreground cleaning. By comparing stacked images and spectra with mock simulations, they demonstrated that the central excess in the stacked signal is robust and consistent with \hi\ emission from the GAMA galaxies. The study also introduced improved covariance estimation based on random shuffling of galaxy positions, allowing systematic effects in the data to be characterised and constrained. Importantly, the stacked spectrum was shown to be sensitive to chromatic systematics such as primary beam ripple \citep{2021MNRAS.502.2970A,Matshawule:2020fjz}, providing a new diagnostic for instrumental effects. This lays the foundation for using stacking not only as a means of measuring average \hi\ emission, but also as a powerful tool for validating calibration and foreground-removal pipelines.


\section{Future prospects and the road ahead}\label{sec:Future}

21cm intensity mapping has the potential to become a valuable resource for cosmology. In the future, traditional galaxy surveys will begin to approach the limits of their reach, constrained by redshift depth and systematic effects that become increasingly dominant as statistical errors shrink. By contrast, 21cm intensity mapping can rapidly chart the three-dimensional structure of the Universe spectroscopically, uniquely reaching high redshifts and motivating next-generation experiments \citep{PUMA:2019jwd,Burns:2021ndk,2023arXiv230110345B,Chen:2024tvn}. The 21cm signal is inherently tied to the astrophysics of hydrogen, its abundance and distribution shaped by processes such as reionisation, star formation, and the subsequent evolution of galaxies that host neutral gas. Fortunately, going to sufficiently \textit{large}, \textit{linear} scales simplifies the impact of these complexities, as the connection between astrophysics and the \hi\ power spectrum enters primarily through the overall amplitude of the clustering\footnote{Plus subdominant shot-noise terms.}, allowing cosmological information to be extracted with \textcolor{black}{fewer degeneracies with the complex} astrophysics \citep{Wolz:2017rlw,Beane:2018pmx,Kovetz:2019uss,Bernal:2019jdo}.

Leveraging the exquisite sensitivity and collecting area of MeerKAT, MeerKLASS is already delivering maps of unprecedented scale, setting the stage for a new era of radio cosmology. The coming years will see exciting developments with its planned 10,000\,deg$^2$ survey spanning $0.4\txteq{<}z\txteq{<}1.45$, providing the largest-volume spectroscopic survey in the southern hemisphere, before the SKAO comes online.

In this section, we introduce some selected forecasts that preview the performance of the full MeerKLASS survey and provide a flavour of its potential. The formalism used in this section is detailed in \appref{app:Fisher} and the Fisher forecast code is publicly available at: \href{https://github.com/meerklass/MeerFish}{\texttt{https://github.com/meerklass/MeerFish}}. We will extend the complexity of these forecasts in follow-up work, but for now, we ensure that all major caveats and assumptions connected with the results are clearly mentioned. For these forecasts, we assume the 2028 MeerKLASS survey (see \autoref{tab:MKobs}), that is a $10{,}000\,\text{deg}^2$ UHF-band survey. We assume the $2{,}500\,$hrs of observations have a conservative 50\% loss due to RFI, hence all used observing times in the forecasts are $1{,}250\,$hrs. We confirm all default survey parameters assumed for MeerKLASS in \autoref{tab:SurveyTable}, along with a next-generation SKAO-Mid Band 1 survey \citep{SKA:2018ckk}, which we also include in some forecasts.

\subsection{Forecasting MeerKLASS potential}

MeerKLASS will primarily seek to measure the \hi\ power spectrum, whose cosmological signal can be modelled as, ignoring redshift dependence for brevity,
\begin{equation}
\begin{split}
    P_\hi(k,\mu) = \alpha_{\|}^{-1} \alpha_{\perp}^{-2}\,\overline{T}^2_\hi\, & \Big(b_\hi + f\mu^2 + b^\hi_\phi f_{\mathrm{NL}} \mathcal{M}(k)^{-1}\Big)^2 P_{\rm m}(k) \\
\end{split}
\end{equation}
There will be observational or instrumental effects that modulate this power spectrum, perhaps most significantly the beam and additive thermal noise, which can be cast into an observer-space power spectrum given by
\begin{equation}\label{eq:P_HI_obs_orig}
    P^\text{obs}_\hi(k,\mu) = P_\hi(k,\mu)\mathcal{B}^2_{\rm beam}(k,\mu) + P_{\rm N}\,.
\end{equation}
We provide the details of these models in \appref{app:Fisher}, explicitly walking through all the terms and discussing how we model the covariance, multipole expansion, and multi-tracer Fisher forecasting.

\begin{table}
    \setlength{\tabcolsep}{4.5pt}
	\centering
	\begin{tabular}{lccc} 
		\toprule
		  & & \textbf{MeerKLASS}  & \textbf{SKAO-MID}\\
		\textbf{Survey Parameters} & & UHF-band & Band 1 \\
		\toprule
        Bandwidth & $\nu_\text{min}$ & 580\,MHz & 350\,MHz\\
         & $\nu_\text{max}$ & 1{,}000\,MHz & 1{,}050\,MHz \\
    	\midrule
        Redshift range & $z_\text{min}$ & 0.4 & 0.35 \\
         & $z_\text{max}$ & 1.45 & 3 \\
		\midrule
        Channel width & $\delta\nu$ & $132.8\,$kHz & $10.8\,$kHz \\
        Redshift resolution (at $z{=}1$) & $\delta z$ & $0.37{\times}10^{-3}$ & $0.030{\times}10^{-3}$ \\
        Survey sky area & $A_\mathrm{sur}$ & 10{,}000\,deg$^2$ & 20{,}000\,deg$^2$\\
        Total observation time & $t_\mathrm{tot}$ & 2{,}500\,hrs & 10{,}000\,hrs \\
        Survey efficiency & $\varepsilon$ & 0.5 & 0.5 \\
        Useable observaton time & $t_\mathrm{obs}$ & 1{,}250\,hrs & 5{,}000\,hrs \\
		Number of dishes & $N_\mathrm{dish}$ & $64$ & $197$ \\
        Dish diameter & $D_\mathrm{dish}$ & $13.5$\,m & $15$\,m\\
        \bottomrule
	\end{tabular}
    \caption{Specifications for the MeerKLASS, UHF-band, and SKAO-MID Band 1 surveys \citep{MeerKLASS:2017vgf,SKA:2018ckk}. 64 of the SKAO dishes will be the existing MeerKAT dishes that have $D_\mathrm{dish}=13.5\,\text{m}$, but for simplicity we make the approximation that all dishes have the same $15\,\text{m}$ diameter.}
    \label{tab:SurveyTable}
\end{table}

\begin{figure}
    \centering
    \includegraphics[width=0.7\linewidth]{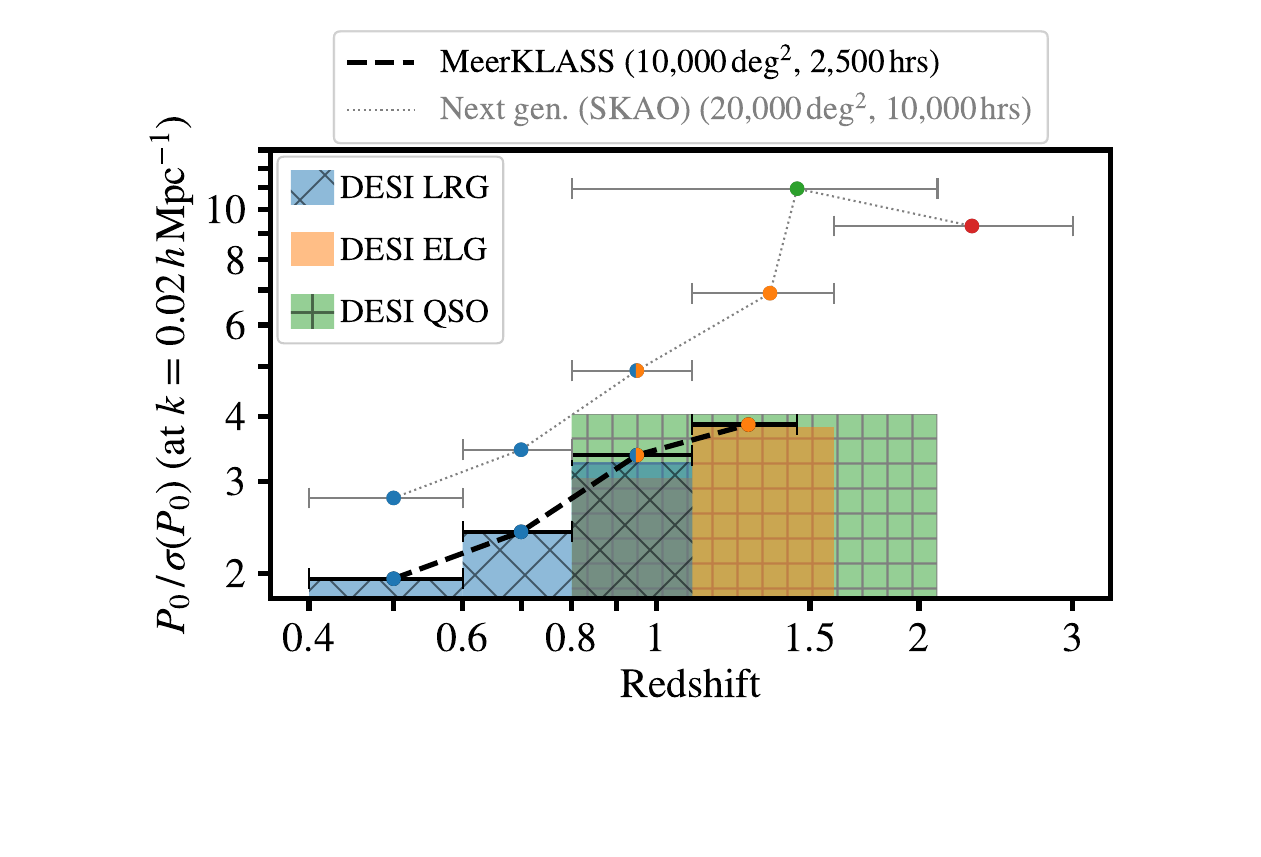}
    \caption{Power spectrum monopole signal-to-noise for MeerKLASS, demonstrating its competitive potential if systematics can be controlled. \textit{Current} state-of-the-art DESI DR2 \citep{DESI:2025zgx} is shown for comparison. The DESI redshift bins are used for the 21cm forecasts (indicated by the coloured points which pair to the coloured bars for each bin), plus one additional distinct high-$z$ bin for the next generation SKAO-Mid Band 1 intensity mapping survey (red point).}
    \label{fig:FisherForecasts}
\end{figure}

\autoref{fig:FisherForecasts} shows the forecast signal-to-noise ($S/N$) on the spherically-averaged (i.e.\ monopole $P_0$) power spectrum measurement at a nominal scale of $k\txteq{=}0.02\hMpc$, suitable for large-scale cosmology where 21cm intensity mapping is most efficient. For simplicity, we assume in all forecasts that signal loss from foreground cleaning has been unbiasedly reconstructed. This is not an unreasonable assumption given the current development of the foreground transfer function \citep{Cunnington:2023jpq,Chen:2025til} (see \secref{sec:TF}), however, it is inevitable that this process will increase uncertainties on large scales (small-$k$), which, in lieu of a reliable covariance model for signal reconstruction, has not been included and we leave this extension to future work.

For comparison, \autoref{fig:FisherForecasts} also shows the $S/N$ from recent DESI data (coloured bars). We take the galaxy catalogue counts and survey areas from DESI DR2 \citep{DESI:2025zgx}, for each of their redshift bins, and compute the mean galaxy number densities, which are used in our modelling formalism to estimate $P/\sigma{P}$, assuming $b_\text{g}\txteq{=}2.0,1.2,2.1$ for the LRG, ELG and QSO samples respectively \citep{DESI:2025qqy}. The intensity maps are matched to these DESI redshift bins where possible (indicated by the coloured points, which pair to each coloured DESI redshift bin). This suggests MeerKLASS can approximately equal the DR2 $S/N$ levels across the available UHF redshift range. We will demonstrate shortly how this can be leveraged for competitive constraints on cosmological parameters. The next generation SKAO $S/N$ is shown by the grey-dotted line, for the same redshift bins as DESI DR2, plus one additional high-redshift bin at $1.6\txteq{<}z\txteq{<}3$ (red point). We note that DESI will continue to gather data, and thus the DESI results presented will improve in terms of statistical precision and systematic control, leading to enhanced signal-to-noise. The 5-year DESI survey began in 2021 and is set to gather 20 million unique redshifts by its completion \citep{DESI:2025fxa,DESI:2024uvr}.

\subsubsection{Optimising survey strategy by increasing sky area} 

Intensity map observation time can be deployed reasonably flexibly across the sky, allowing MeerKLASS to determine the optimal balance between depth and area. This flexibility enables strategic choices in how to maximise the cosmological return, guided by forecast analyses that quantify the trade-offs.  

These forecasts consistently show a preference for a ``wide and shallow'' survey strategy i.e.\ spreading available observation time over larger areas with fewer repeated scans. The reason this approach is favoured is that, up to a certain limit, cosmic variance dominates the error budget over thermal noise. Whilst spreading the total observing time over a larger footprint increases the pixel noise floor due to shorter integration per pointing (see \appref{app:Fisher}, particularly \autoref{eq:P_N}), the reduction in cosmic variance more than compensates. This leads to smaller overall uncertainties on cosmological parameters compared with a ``narrow and deep'' strategy. In other words, wider surveys deliver more independent Fourier modes, and this outweighs the higher instrumental noise per pixel.  

\begin{figure}
    \centering
    \includegraphics[width=0.5\linewidth]{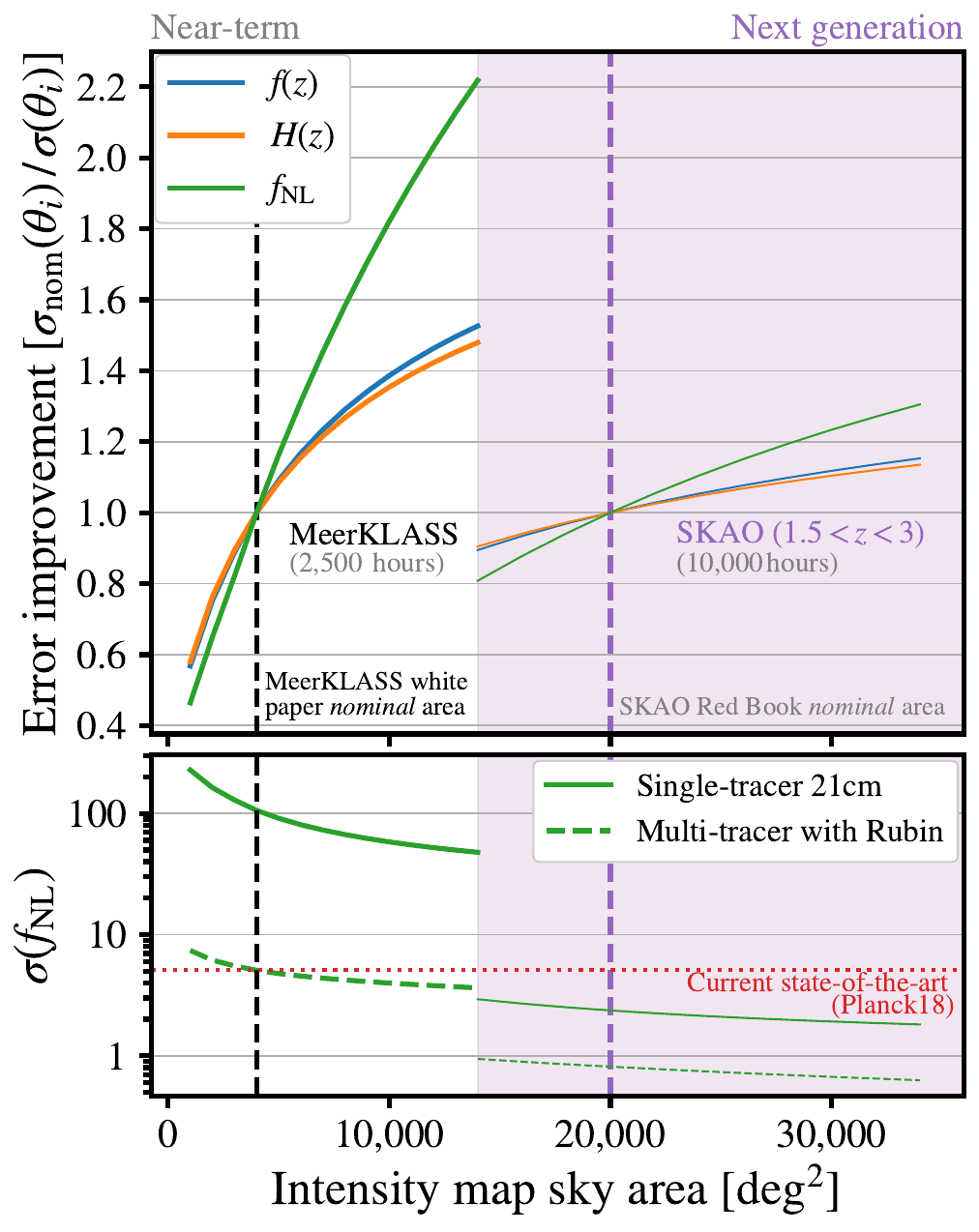}
    \caption{Forecast improvement on constraints by changing sky area for a \textit{fixed} observational time (quoted in grey). For \fNL, $\txteq{>}100$\% gains on the original proposed \cite[nominal][]{MeerKLASS:2017vgf} area are available by adopting a ``\textit{wide and shallow}'' MeerKLASS survey. Bottom panel shows the absolute error on \fNL. \textcolor{black}{A multi-tracer approach with a Rubin Observatory galaxy clustering probe (green-dashed line) will see Planck18 surpassed}, and primordial non-Gaussianity detections become viable. Next generation potential is shown in purple shading \cite[following][]{SKA:2018ckk}. \textcolor{black}{We assume the full Y10 gold sample for Rubin with the SKAO, but a restricted version for MeerKLASS (details given in the text).}}
    \label{fig:params_vs_Area}
\end{figure}

This behaviour is illustrated in the top panel of \autoref{fig:params_vs_Area}, which shows how the forecast errors on cosmological parameters ($f(z)$, $H(z)$, and $f_{\rm NL}$), from the \hi\ power spectrum measurement improve as the survey area is expanded. In each case, the constraints tighten as the footprint increases, highlighting the statistical power of wide-area coverage. In particular, the gains for $f_{\rm NL}$ are striking (discussed further in the upcoming \secref{sec:fNL}), reflecting its sensitivity to ultra-large-scale modes that can only be accessed by surveying the largest possible volumes.  

In this simple test we let the parameters $\boldsymbol{\theta}\txteq{=}\{b_\hi,f,\alpha_\perp,\alpha_\parallel,f_\text{NL}\}$ vary, whilst assuming everything else fixed. We extract $H(z)$ from the AP parameter $\alpha_\parallel(z)$ (see \autoref{eq:APparams}), and marginalise over $b_\hi$ and $\alpha_\perp$. The parameters $b_\hi$ and $f$ should be fully degenerate in linear theory for the monopole of the \hi\ power spectrum. For intensity mapping, it has been shown that going to higher order multipoles is especially crucial for breaking such degeneracies \citep{Bernal:2019jdo,Kennedy:2021srz,Rubiola:2021afc}, so in this analysis we include the quadrupole and hexadecapole, i.e. $\boldsymbol{P}_\ell\txteq{=}\{P_0,P_2,P_4\}$, which assumes observational anisotropies can be well controlled \citep{Blake:2019ddd,Cunnington:2020mnn,Soares:2020zaq}. There will be additional degeneracies with $\overline{T}_\hi$ and $\sigma_8$, for which we fix here, thus we are inherently assuming they have been independently well measured, or alternatively constraints would include these factors as degenerate quantities, e.g. $f\sigma_8\overline{T}_\hi$. The $\sigma_8$ degeneracy is a general problem for large-scale structure, but the $\overline{T}_\hi$ is, of course, more unique to intensity mapping \cite[identified in ][]{Bernal:2019jdo,Castorina:2019zho}. Exploiting \textit{tomography}, i.e. \textcolor{black}{leveraging} the differing redshift evolution of degenerate parameters across bins, can provide some immediate relief here. Additionally, 1-point statistics, such as the voxel intensity distribution (VID) \citep{Breysse:2016szq,Sato-Polito:2022fkd,Bernal:2023ovz}, can capture the non-Gaussian information in the line intensity maps, thus combining the VID with the power spectrum helps to break degeneracies between astrophysical and cosmological parameters \citep{Sabla:2024dxz}. Furthermore, going to non-linear scales with interferometers, where scale dependence will enter and since \hi\ clustering should be accurately described by perturbative methods at these mildly non-linear scales, the $\overline{T}_\hi$ degeneracy can potentially be broken \citep{Castorina:2019zho,Pourtsidou:2022gsb}. We defer an extended study on this topic involving simulations to future work.

\autoref{fig:params_vs_Area} motivates the decision for MeerKLASS to pursue the much larger $10{,}000\degsq$ survey (as is now planned) compared to the proposed $4{,}000\degsq$ in the original white paper \citep{MeerKLASS:2017vgf}. Nevertheless, there are important caveats to adopting a wide-and-thin strategy. First, additive systematics are less effectively averaged down in shallower data, making the survey more vulnerable to residual foregrounds/RFI and instrumental artefacts \citep{Li:2020bcr,Wang:2020lkn,Engelbrecht:2024eoc,Chen:2025bcx}. Second, wider coverage will inevitably push observations closer to the Galactic plane, where foregrounds are brighter and more complex \citep{Staveley-Smith:2016gst,2024MNRAS.531..649G,Irfan:2021xuk}. This raises the requirement for robust foreground-cleaning methods to perform reliably at lower Galactic latitudes. Both issues will be investigated in detail using ongoing MeerKLASS observations, which will provide an empirical testbed for evaluating the performance of foreground cleaning and systematic control across different regions of sky.  

\subsection{Probing the growth rate of structure at radio wavelengths: independent tests of gravity}

\begin{figure}
    \centering
    \includegraphics[width=0.6\linewidth]{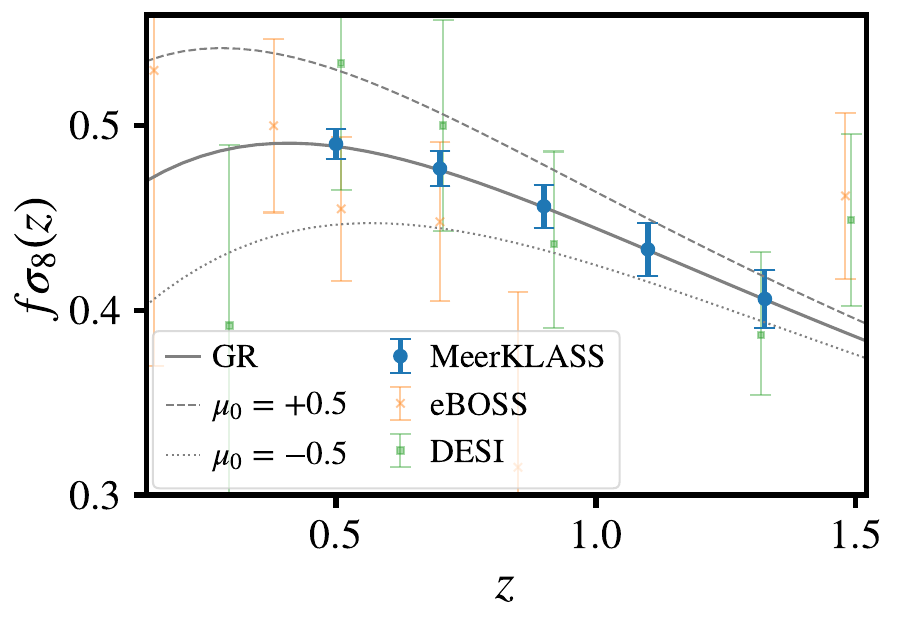}
    \caption{Forecast constraints on growth rate from full MeerKLASS $10{,}000\degsq$ survey, showing its potential to distinguish general relativity (GR) from modified theories of gravity; parameterised by $\mu_0$. Constraints are compared to the current state-of-the-art from eBOSS and DESI \citep{eBOSS:2020yzd,DESI:2024jxi}.}
    \label{fig:growth_rates}
\end{figure}

Redshift-space distortions (RSD) in the large-scale distribution of matter provide a direct probe into the growth rate of structure, $f(z)$ \citep{Huterer:2022dds}. Measuring $f(z)$ across redshift enables stringent tests of gravity, since any departure from the predictions of General Relativity would manifest as a change in the growth history. In this way, RSD measurements complement distance-scale probes such as baryon acoustic oscillations (BAO), offering a route to test whether extensions to Einstein’s theory are required \citep{Joyce:2014kja}.  

Unlike BAO, which are highly robust to observational systematics \citep{eBOSS:2021rwq}, RSD measurements are more vulnerable to survey-specific effects \citep{Samushia:2011cs} that can introduce anisotropies in the clustering signal. This increases the importance of obtaining independent RSD measurements across different experiments, where uncorrelated systematics can be identified and mitigated in combined analyses. Intensity mapping offers such an opportunity: by probing the same underlying large-scale modes with a different set of contaminants and instrumental systematics, MeerKLASS can provide an independent validation of growth rate measurements.  

\autoref{fig:growth_rates} illustrates the forecast performance of MeerKLASS in constraining the growth rate. The plot shows how intensity mapping can act as a powerful discriminator between modified gravity models, here parameterised by $\mu_0$, which governs deviations from GR in the effective Newtonian coupling \citep{Simpson:2012ra,DES:2018ufa}. MeerKLASS is forecast to deliver competitive constraints, exceeding current state-of-the-art spectroscopic surveys such as eBOSS \citep{eBOSS:2020yzd} and DESI \citep{DESI:2024jxi}, and extending measurements to complementary redshift ranges. This demonstrates the potential of MeerKLASS as an effective “ruler” on modified gravity.  

We make no attempt to include any non-linear effects which would typically reduce sensitivity for RSD probes in a more conventional optical galaxy survey analysis, which can make measurements out to very small scales. However, for single-dish 21cm intensity mapping with MeerKLASS, the effect of the large beam reduces the contribution from small scales massively. Therefore, not accounting for non-linear modelling challenges in this forecast is unlikely to be a large source of oversight. We can demonstrate this with a simple test in which we compare the growth constraints from two cases: \textit{i}) including non-linear scales, up to $k\txteq{<}0.3\hMpc$, and \textit{ii}) with a more restricted range that excludes most non-linear scales with $k\txteq{<}0.15\hMpc$. Doing this in the lowest redshift bin in \autoref{fig:growth_rates}, where the beam is smallest and non-linear effects will be largest, still only causes a 5\% reduction in the constraint. Thus providing evidence that we are reasonably immune to non-linear complications.

The main challenge, however, will be to demonstrate sufficient control of observational systematics. Foreground cleaning, beam chromaticity, and survey masks all introduce anisotropies that can mimic or bias an RSD signal if not modelled with high fidelity \citep{Cunnington:2020mnn}. Developing robust pipelines that accurately capture these effects will therefore be critical. Future work with MeerKLASS will focus on meeting this challenge, with the goal of contributing a new, independent constraint on the growth rate of structure that will strengthen the global programme of testing gravity with large-scale structure.  


\subsection{Multi-tracer cosmology with MeerKLASS for primordial non-Gaussianity (PNG)}\label{sec:fNL}

Perhaps the most powerful opportunity offered by intensity mapping is its spectroscopic access to ultra-large scales, where few other probes can compete \citep{Alonso:2015uua}. On these scales, the 21cm line provides a unique window into the physics of the early Universe, particularly primordial non-Gaussianity (PNG) \citep{Komatsu:2001rj,Dalal:2007cu}. PNG refers to deviations from Gaussian initial conditions seeded during inflation, which imprint themselves as a scale-dependent modulation of galaxy and \hi\ bias on the largest modes and is modulated by the parameter $f_{\rm NL}$. Measuring this so-called \textit{local}-$f_{\rm NL}$ with high precision is one of the most linear methods, thus less complicated by small-scale astrophysics, for distinguishing between competing models of inflation.  

Access to the largest scales is therefore crucial for optimal PNG constraints. The bottom panel of \autoref{fig:params_vs_Area} demonstrates this point, showing the forecast $\sigma(f_{\rm NL})$ as a function of sky area. Due to our adoption of the realistic simulation-based PNG bias for \hi\ \cite[see][and Appendix \autoref{eq:HIPNGbias}]{Barreira:2021dpt}, the single-tracer constraint with MeerKLASS does not reach particularly competitive \fNL\ values\footnote{This improves to $\sigma(f_{\rm NL})\txteq{=}15.1$ if adopting the standard universality value for $b_\phi^\hi$.} However, impressive gains are achieved through a multi-tracer approach \citep{Seljak:2008xr,Fonseca:2015laa,Squarotti:2023nzy}. By cross-correlating MeerKLASS with an overlapping survey that has a sufficiently different bias, cosmic variance can be reduced between the tracers, dramatically reducing the statistical error on $f_{\rm NL}$ \cite[see e.g.][for an early forecast of MeerKAT${\times}$DES]{Fonseca:2016xvi}. The forecast for a joint analysis with the Rubin Observatory \citep{LSSTScience:2009jmu} is shown in \autoref{fig:params_vs_Area} (dashed green line), and at $10{,}000\degsq$ reaches $\sigma(f_{\rm NL})\txteq{=}3.27$ if assuming the \textcolor{black}{simulation-derived TNG300} $b_\phi^\hi$ model from \citep{Barreira:2021dpt}, \textcolor{black}{(see discussion surrounding \autoref{eq:HIPNGbias} for more details on the primordial bias parameter)}. Rubin, a forthcoming deep optical imaging survey, offers an ideal multi-tracer companion thanks to its wide area and potentially complete southern sky overlap with MeerKLASS. For Rubin, we assume a Y10 gold sample number density of 48\,arcmin$^{-2}$ \citep{LSSTDarkEnergyScience:2018jkl} for the next generation constraints, over a $18{,}000\,$deg$^2$ sky area. For the earlier MeerKLASS constraints, we assume Rubin has a smaller area of $10{,}000\,$deg$^2$ with perfect MeerKLASS overlap and approximately half its final number density. We assume a photometric redshift error of $\sigma_z\txteq{=}0.02(1\txteq{+}z)$ for all Rubin surveys \citep{LSSTScience:2009jmu}, although this has only a mild impact for the large-scale \fNL\ probe,  ${\sim}\,11$\% (${\sim}\,6$\% for SKAO) increase in \fNL\ error compared to perfect redshift precision (i.e. $\sigma_z\txteq{=}0$). For the Rubin galaxy bias, we assume the simple model of $b_\text{g}\txteq{=}1\txteq{+}0.84z$, following \citep{LSSTScience:2009jmu}.

The combination of MeerKLASS and early-Rubin is therefore forecast to provide competitive $\sigma(f_{\rm NL})$ values, potentially surpassing the current state-of-the-art CMB constraints from Planck \citep{Planck:2018vyg}. Looking ahead, the synergy between the full SKAO survey and Rubin could push uncertainties below the crucial threshold of $\sigma(f_{\rm NL}) < 1$, a milestone that would allow many multi-field inflationary models to be definitively ruled out. Achieving this level of precision would represent a landmark result in fundamental cosmology.  

Realising this potential will require stringent control of observational systematics. Foreground cleaning and the associated problem of signal loss are especially critical, since they predominantly affect the large-scale modes that drive PNG sensitivity. However, as shown in \citet{Cunnington:2023jpq}, robust reconstruction of the transfer function offers a viable pathway even for precision measurements of this kind. The incoming MeerKLASS data therefore provide an ideal test-bed for developing and validating these methods, paving the way towards next-generation multi-tracer PNG cosmology.  

Additional challenges for local \fNL\ constraints, more general to large-scale structure probes, is the degeneracy with the primordial bias parameter $b_\phi$. Detections of $f_\text{NL}\txteq{\neq}0$ are thus reliant on a tight prior in the quantity $b_\phi f_\text{NL}$. As with galaxy surveys, the value of $b_\phi$ is not currently well constrained for the \hi\ tracer \citep{Barreira:2021dpt,Barreira:2022sey}. Future work using the latest state-of-the-art hydrodynamical simulations \cite[e.g.][]{Schaye:2025xuv} can explore this issue further. Our multi-tracer approach will bring benefits to this problem too by reducing the sensitivity to each tracer’s $b_\phi$ prior
uncertainty \citep{Barreira:2023rxn}, especially when tracers differ in redshift distribution and bias evolution, exactly the case for MeerKLASS and Rubin.

\section{Summary}\label{sec:Summary}

MeerKLASS has established itself as a pioneering programme for validating and exploiting single-dish 21cm intensity mapping with the MeerKAT array. Through careful development of calibration, map-making, and analysis methods, the collaboration has overcome key challenges including foreground contamination, signal loss, and the transformation of survey maps into Cartesian volumes for Fourier analysis. These efforts have culminated in the first robust detections of cosmological clustering in cross-correlation with optical galaxy surveys, alongside significant progress towards recovering auto-correlation \textcolor{black}{power spectra} measurements and stacked \hi\ emission.

Building on these early successes, forecasts demonstrate the powerful potential of MeerKLASS to deliver competitive cosmological constraints across a broad redshift range, particularly when its large-area UHF-band survey is completed. Its measurements of the growth rate of structure, baryon acoustic oscillations, and primordial non-Gaussianity will provide independent and complementary probes to optical surveys, while also serving as a crucial pathfinder for the SKAO.

While significant challenges remain, especially the control of systematics and the robust modelling of observational effects, the incoming data stream represents a transformative opportunity. MeerKLASS is now positioned to open a new window for cosmology, establishing 21cm intensity mapping as a unique and powerful probe of the large-scale Universe.

\section*{Statements \& Declarations}

This paper was prepared for a contribution to the Springer Nature Astronomy Prize Awardees Collections (2025). SCu would like to thank Springer Nature for this invitation and also the Royal Astronomical Society for the 2025 Early Career Award.

\subsubsection*{Funding}

SCu acknowledges support from the UKRI Stephen Hawking Fellowship (grant reference EP/U536751/1). 
MBS and SCa acknowledge support from the Italian Ministry of Foreign Affairs and International
Cooperation (\textsc{maeci}), Grant No.\ ZA23GR03. SCa also acknowledges support, from the Italian Ministry of University and Research (\textsc{mur}), PRIN 2022 `EXSKALIBUR – Euclid-Cross-SKA: Likelihood Inference Building for Universe's Research', Grant No.\ 20222BBYB9, CUP D53D2300252 0006, and from the European Union -- Next Generation EU.
JLB acknowledges funding from the project UC-LIME (PID2022-140670NA-I00), financed by MCIN/AEI/ 10.13039/501100011033/FEDER, UE.
IPC is supported by the European Union within the Next Generation EU programme [PNRR-4-2-1.2 project no.\ SOE\textunderscore0000136, RadioGaGa]. 
JF acknowledges support of Funda\c{c}\~{a}o para a Ci\^{e}ncia e a Tecnologia through the Investigador FCT Contract no.\ 2020.02633.CEECIND/CP1631/CT0002, the FCT exploratory project 10.54499/2023.15069.PEX, and the research grant UID/04434/2025.
MGS acknowledges support from the South African National Research Foundation (Grant No. 84156). 
JW acknowledges support from the National SKA Program of China (no.\ 2020SKA0110100) and the National Natural Science Foundation of China (NSFC, Grant No. 12573112).

\subsubsection*{Competing interests}

The authors have no relevant financial or non-financial interests to disclose.

\subsubsection*{Ethics declaration} Not applicable
%









\begin{appendices}

\section{Fisher forecast formalism}\label{app:Fisher}

We begin by reintroducing the \hi\ power spectrum as in \secref{sec:Future} but now explicitly stating redshift dependence, and casting in terms of fiducial scales $k^\text{f}$ and $\mu^\text{f}$ (which we explain shortly)
\begin{equation}\label{eq:P_HI_sig}
\begin{split}
    P_\hi(k^\text{f},\mu^\text{f},z) = \alpha_{\|}^{-1}(z) \alpha_{\perp}^{-2}(z)\, & \overline{T}^2_\hi(z)\,\Big(b_\hi(z) + f(z)\mu^2 \\
    & + b^\hi_\phi(z) f_{\mathrm{NL}} \mathcal{M}(k, z)^{-1}\Big)^2 P_{\rm m}(k, z) \,.
    \end{split}
\end{equation}
To parameterise the normalisation of the matter power spectrum, it is conventional to use $P_\text{m}\txteq{=}\sigma_8P_\text{m,8}$, where $\sigma_8$ is RMS of the density fluctuations within a sphere of radius $8\hMpc$, such that
\begin{equation}
\begin{split}
    P_\hi(k^\text{f},\mu^\text{f},z) = \alpha_{\|}^{-1}(z) \alpha_{\perp}^{-2}(z)\, & \overline{T}^2_\hi(z)\,\Big(b_\hi(z)\sigma_8(z) + f(z)\sigma_8(z)\mu^2 \\
    & + b^\hi_\phi(z) f_{\mathrm{NL}}\sigma_8(z) \mathcal{M}(k, z)^{-1}\Big)^2 P_{\rm m,8}(k, z)\,,
\end{split}
\end{equation}
so that the $\sigma_8$ parameter makes the degeneracy between parameters and power spectrum amplitude explicit. In \autoref{eq:P_HI_sig}, $\overline{T}_\hi$ is the background mean \hi\ brightness temperature, $b_\hi$ is the linear bias for \hi, and $f$ is the growth rate of structure. For the \hi\ bias,  we fit to hydrodynamical simulations \citep{Villaescusa-Navarro:2018vsg} providing the model
\begin{equation}
    b_\hi(z) = 0.842 + 0.693z - 0.0459z^2\,.
\end{equation}
The mean \hi\ brightness temperature is modelled using 
\begin{equation}
    \overline{T}_\hi(z) = 180\, \Omega_\hi \frac{ h \, (1+z)^2}{ H(z)/H_0}\,\text{mK}\,,
\end{equation}
where we use the model for $\Omega_\hi$ based on the one used in \citet{SKA:2018ckk} but adapted based on the latest $\Omega_\hi$ constraints at higher redshift from MeerKLASS intensity mapping \citep{Cunnington:2022uzo}
\begin{equation}
    \Omega_\hi(z) = 0.00067432 + 0.00039z - 0.000065z^2\,.
\end{equation}
In \autoref{eq:P_HI_sig}, we include the possibility for local primordial non-Gaussianity (PNG) to influence the power, which occurs by the bias acquiring a scale-dependent correction given by the $b_\phi f_\text{NL}\mathcal{M}^{-1}$ term. $f_\text{NL}$ modulates the PNG contribution and we assume $f_\text{NL}\txteq{=}0$ as fiducial default. $b^\hi_\phi$ is the PNG bias parameter for \hi, where is common to adopt the standard universality assumption of the halo mass function (which indeed we use for the galaxy power spectrum). However, to be more realistic, we instead interpolate the PNG bias for \hi\ from the simulation results in \citet{Barreira:2021dpt}, using the model
\begin{equation}\label{eq:HIPNGbias}
    b^\hi_\phi(z) = -1.67 + 2.40z - 0.19z^2
\end{equation}
The final factor relating to PNG is given by
\begin{equation}
    \mathcal{M}(k,z)\txteq{=}(2/3) k^2 T_\text{m}(k) D_{\mathrm{md}}(z) /(\Omega_\text{m} H_0^2)\,,
\end{equation}
where $T_\text{m}$ is the matter transfer function and $D_\text{md}$ is the linear growth factor normalized to $1/(1\txteq{+}z)$. 

21cm intensity mapping pixel positions in redshift and angular sky coordinates are measured directly, but converting them into comoving distances requires assuming a fiducial cosmology. If this assumed model differs from the true one, the inferred radial and transverse scales are distorted differently, introducing an artificial anisotropy in the power spectrum. This purely geometric distortion is known as the Alcock–Paczynski (AP) effect \citep{Alcock:1979mp}. The effect can be described by the rescaling factors present in \autoref{eq:P_HI_sig}
\begin{equation}\label{eq:APparams}
    \alpha_{\perp}(z)=\frac{D_\text{A}(z) / r_{\mathrm{s}}}{\left(D_\text{A}(z) / r_{\mathrm{s}}\right)_{\mathrm{f}}}\,, \quad \alpha_{\|}(z)=\frac{\left(H(z) r_{\mathrm{s}}\right)_{\mathrm{f}}}{H(z) r_{\mathrm{s}}}\,.
\end{equation}
and the true wavenumbers are distorted by
\begin{equation}
    k_\perp = k_\perp^\text{f}/\alpha_\perp\,, \quad k_\parallel = k_\parallel^\text{f}/\alpha_\parallel\,.
\end{equation}
Here the fiducial indices denote the fiducial (reference) cosmology assumed in the \textit{measured} wavenumbers, and the wavenumber without an $\text{f}$ index denotes the \textit{true} wavenumbers for the true cosmology. These are conventionally transposed into $k$ and $\mu$ using the factor $F_\text{AP}\txteq{=}\alpha_\parallel/\alpha_\perp$, where
\begin{equation}
    k=\frac{k_{\mathrm{f}}}{\alpha_{\perp}}\left[1+\left(\mu_{\mathrm{f}}\right)^2\left(F_{\mathrm{AP}}^{-2}-1\right)\right]^{1 / 2}
\end{equation}
\begin{equation}
    \mu=\frac{\mu_{\mathrm{f}}}{F_{\mathrm{AP}}}\left[1+\left(\mu_{\mathrm{f}}\right)^2\left(F_{\mathrm{AP}}^{-2}-1\right)\right]^{-1 / 2} .
\end{equation}
To correct for any change in volume caused by the different cosmologies, the power spectrum model must be multiplied by $\alpha_\parallel^{-1}\alpha_\perp^{-2}$, which explains this factor at the front of \autoref{eq:P_HI_sig}. Lastly, $P_\text{m}$ in \autoref{eq:P_HI_sig}, is the matter power spectrum obtained from a Boltzmann solver \cite[we used \texttt{CAMB}][as standard]{Lewis:1999bs}.

Similarly to the \hi\ power spectrum, we can generically model the \textit{galaxy} power spectrum as
\begin{equation}\label{eq:P_g_sig}
    P_{\rm g}(k^\text{f},\mu^\text{f},z) = \alpha_{\|}^{-1}(z)\, \alpha_{\perp}^{-2}(z)\,\Big(b_{\rm g}(z) + f(z)\mu^2 + b^{\rm g}_\phi(z) f_{\mathrm{NL}} \mathcal{M}(k, z)^{-1}\Big)^2 P_{\rm m}(k, z)\,,
\end{equation}
where we assume $b_\text{g}\txteq{=}1\txteq{+}0.84z$, for a Rubin-like bias \citep{LSSTScience:2009jmu}, and $b_\text{g}\txteq{=}2.0,1.2,2.1$ for the DESI LRG, ELG and QSO samples respectively. We adopt the standard universality assumption for the \textit{galaxy} PNG bias 
\begin{equation}\label{eq:PNG_universality}
    b^\text{g}_\phi(z) = 2\,\delta_\text{c} \left(b_\text{g}(z) - 1\right)\,,
\end{equation}
where $\delta_\text{c}\txteq{=}1.686$. 

\medskip
\noindent
\textbf{Observable power spectrum}:
The pure cosmological power spectrum in \autoref{eq:P_HI_sig} is not what is measured in practice. The observed fluctuations in the intensity maps (or galaxy field) are modulated by instrumental effects and contain additive noise. The delicate handling of such effects is far from trivial, and we discuss some of the subtleties of this in \appref{sec:ObsImpact}. We use a simple observer-space \hi\ power spectrum modelled by
\begin{equation}\label{eq:P_HI_obs}
    P^{\rm obs}_\hi(k^\text{f},\mu^\text{f},z) = P_\hi(k^\text{f},\mu^\text{f},z)\,\mathcal{B}^2_{\rm beam}(k^\text{f},\mu^\text{f},z) + P_{\rm N}(z)\,,
\end{equation}
which includes impact from two main instrumental effects coming from additive thermal noise (introduced by the noise power $P_\text{N}$) and the telescope beam profile which acts as a convolution on the \hi\ power determined by the window $\mathcal{B}_\text{beam}$, which damps fluctuations in the transverse directions. For the beam damping, assuming a perfectly Gaussian-shaped profile, its window is given by,
\begin{equation}
    \mathcal{B}_{\rm beam}(k,\mu,z) = \exp \left[-\frac{1}{2} k_{\perp}^2 R_{\mathrm{b}}^2(z)\right]=\exp \left[-\frac{k^2}{2}(1-\mu^2) R_{\mathrm{b}}^2(z)\right]\,,
\end{equation}
where $R_\text{b}(z)\,{=}\,\chi(z)\,\theta_\text{FWHM}(z)/2\sqrt{2\ln 2}$. $\theta_\text{FWHM}$ is radio telescope specific but for a generic dish size with diameter $D_\text{dish}$ and observations at a frequency $\nu$, we can say $\theta_\text{FWHM}\txteq{=} c\,/\,(\nu D_\text{dish})$. Within a Fisher forecast or simulation, this beam term is known exactly, and if desired, can be perfectly deconvolved from the power spectrum (as we explore in \appref{sec:ObsImpact}). In real analyses, however, the beam can be more complex and unknown; hence, it would be impractical to deconvolve fully from the data. We therefore leave the beam in the signal and model it to more closely match a real data analysis approach \cite[as discussed in][]{Bernal:2019jdo}.

The noise power in \autoref{eq:P_HI_obs} is given just by the telescope's thermal noise, i.e. we neglect shot-noise, which will be minimal for a \hi\ intensity mapping survey \citep{Villaescusa-Navarro:2018vsg,Spinelli:2019smg}, hence this is modelled as 
\begin{equation}\label{eq:P_N}
    P_\text{N} = V_\text{pix}\sigma_\text{N}^2
\end{equation}
where the pixel volume is determined by the pixel area $A_\text{pix}$ so that
\begin{equation}
    V_{\mathrm{pix}}=\Omega_{\mathrm{A}} \int_{z_0}^{z_1} \text{d} z \frac{\text{d}V}{\text{d}z \text{d} \Omega}=\Omega_{\mathrm{A}} \int_{z_0}^{z_1} \text{d} z \frac{c \, \chi^2(z)}{H(z)}\,,
\end{equation}
where $\Omega_\text{A} \txteq{=} A_\text{pix} (\pi/180)^2$ and the redshift intervals are determined by the frequency of observation $\nu$ and the telescope's frequency resolution $\delta\nu$ so that $z_0=\nu_\text{21cm}/(\nu{+}\delta\nu)\txteq{-}1$ and $z_1=\nu_\text{21cm}/(\nu{-}\delta\nu)\txteq{-}1$, with $\nu_\text{21cm}\txteq{=}1420.4\,$MHz, the frequency of the 21cm line. The RMS of the noise fluctuations is given by the radiometer equation;
\begin{equation}\label{eq:sigma_N}
    \sigma_{\mathrm{N}}(\nu)=\frac{T_{\mathrm{sys}}(\nu)}{\sqrt{2\, \delta\nu \, t_{\mathrm{pix}}}}\,.
\end{equation}
The system temperature is given by
\begin{equation}
    T_{\mathrm{sys}}(\nu)=T_{\mathrm{rx}}(\nu)+T_{\mathrm{spl}}+T_{\mathrm{CMB}}+T_{\mathrm{gal}}(\nu)
\end{equation}
where the contribution from spill-over is $T_\text{spl}\txteq{=}3\,$K, the background contribution from the CMB
is $T_\text{CMB}\txteq{=} 2.73\,$K and the contribution from our own Galaxy is $T_\text{gal}\txteq{=}15\,\mathrm{K}(408\mathrm{MHz} / \nu)^{2.75}$, which is tuned to the match the average sky temperature excluding $|b|\txteq{<}10^\circ$ in galactic latitude. Lastly, the receiver temperature is
\begin{equation}
    T_{\mathrm{rx}}(\nu)=7.5\,\mathrm{K}+10\,\mathrm{K}\left(\frac{\nu}{\mathrm{GHz}}-0.75\right)^2\,,
\end{equation}
which follows \citet{Cunnington:2022ryj} which tuned this model to match recent intensity mapping observations with the MeerKAT pilot survey \citep{Wang:2020lkn}. The time per pixel $t_\text{pix}$ in the noise RMS (\autoref{eq:sigma_N}), is given by the total observation time $t_\text{obs}$ from the survey, shared among each pixel. Thus we define it as 
\begin{equation}
    t_{\mathrm{pix}}=N_{\text{dish}} t_{\text{obs }}\left(\theta_{\mathrm{FWHM}} / 3\right)^2 / A_{\text{sur}}\,
\end{equation}
where we assume each pixel is 1/3 the size of the telescope's beam size (approximately correct for MeerKAT map-making). In single-dish mode, we get a set of observations per dish, hence observation time per pixel is multiplied by the number of dishes $N_\text{dish}$.

The observed counterpart for the galaxies will contain shot-noise as the noise contribution, which is caused by the discreteness of the galaxies and is the power present in the absence of clustering. The observed galaxy power is thus given by
\begin{equation}\label{eq:P_g_obs}
    P^{\rm obs}_{\rm g}(k^\text{f},\mu^\text{f},z) = P_{\rm g}(k^\text{f},\mu^\text{f},z) + \frac{P_{\rm SN}(z)}{\mathcal{B}^2_z(k^\text{f},\mu^\text{f},z)} \,,
\end{equation}
where $P_{\rm SN}\txteq{=}1\,/\,\bar{n}_{\rm g}$ is the galaxy shot noise, with $\bar{n}_{\rm g}$ as the galaxy number density for the survey. Any additional damping factors, such as redshift uncertainty (given by $\mathcal{B}_z$), should not be directly applied to the signal power $P_\text{g}$, again to avoid extracting information from these terms, which are designed to represent loss of information. The redshift uncertainty window is defined as 
\begin{equation}
    \mathcal{B}_z(k^\text{f},\mu^\text{f},z) = \exp \left[-\frac{k^2\mu^2}{2} \sigma_\parallel^2(z)\right]\,,
\end{equation}
where $\sigma_\parallel$ is the spatial radial error, modulated by the telescope's redshift (RMS) error $\sigma_z$, which is determined by $\sigma_\parallel(z)\txteq{=}[c / H(z)] \sigma_z$.

The errors on the power spectrum (\hi\ or galaxy) can be generically estimated as
\begin{equation}\label{eq:sig_err}
    \sigma^2(k, \mu)=\frac{P_\text{obs}^2(k,\mu)}{N_{\operatorname{modes}}(k, \mu)}\,,
\end{equation}
in which we are inherently assuming there is no $k$-mode coupling i.e. a Gaussian covariance approximation. For future 21cm intensity mapping surveys this should become a more robust assumption, but for current results this appears not to be true \citep{MeerKLASS:2024ypg,Chen:2025til}. We leave an exploration into the consequences of wide-spread mode coupling for future work. The number of \textit{unique} modes is given by
\begin{equation}
    N_{\text {modes }}(k,\mu) = \frac{k^2 \Delta k \Delta \mu}{8 \pi^2} V_{\text{sur}}\,,
\end{equation}
where survey volume is determined by the minimum and maximum redshift ($z_{\min}$, $z_{\max}$) of the survey and its area $A_\text{sur}$;
\begin{equation}
    V_{\mathrm{sur}} = \Omega_{\mathrm{sur}} \int_{z_{\min }}^{z_{\max }} \text{d} z \frac{c \, \chi^2(z)}{H(z)}\,,
\end{equation}
where $\Omega_\text{sur}\txteq{=}A_\text{sur} (\pi/180)^2$.

\medskip
\noindent
\textbf{Multipole expansion}: It is convenient to adopt the multipole expansion of the anisotropic power spectrum in \autoref{eq:P_HI_sig} \citep{Cunnington:2020mnn,Bernal:2019jdo,Soares:2020zaq,Berti:2022ilk}, which will be more representative of the estimators applied to observed data. This expansion is achieved using
\begin{equation}
    P(k, \mu)=\sum_{\ell} P_{\ell}(k) \mathcal{L}_{\ell}(\mu)\,,
\end{equation}
\begin{equation}\label{eq:P_ell}
    P_{\ell}(k,z) = (2\ell+1) \int_{0}^1 \mathrm{~d} \mu \, P(k,\mu,z) \mathcal{L}_{\ell}(\mu) \,,
\end{equation}
where
\begin{equation}
    \mathcal{L}_0=1, \quad \mathcal{L}_2=\frac{3 \mu^2-1}{2}, \quad \mathcal{L}_4=\frac{35 \mu^4-30 \mu^2+3}{8}\,.
\end{equation}
and the errors on these multipoles is given by
\begin{equation}
    \sigma_{P_{\ell}}(k)=\sqrt{(2 \ell+1)^2 \int_{0}^1 \mathrm{~d} \mu \, \sigma^2(k, \mu) \mathcal{L}_{\ell}^2(\mu)}\,.
\end{equation}
Note, in other literature, e.g. \citet{Bernal:2019jdo}, which we otherwise follow closely, the AP pre-factor $\alpha^{-1}_\parallel\alpha^{-2}_\perp$ is taken outside the integral of \autoref{eq:P_ell}, but these approaches are identical since the AP parameters are scale-independent and can be taken outside as constants at fixed $z$.

\medskip
\noindent
\textbf{Fisher forecasting}: To include our observable in a Fisher forecast, we consider that the set of fluctuations $\delta(\boldsymbol{x})$ within a survey map's pixels with position $\boldsymbol{x}$ can be reasonably modelled as a Gaussian distribution with mean of zero and pixel covariance $\textbf{\textsf{C}}\txteq{\equiv}\langle\delta^*(\boldsymbol{x})\delta(\boldsymbol{x})\rangle$. The Fisher information matrix is then given by \citep{Tegmark:1997rp,Seo:2003pu,Seo:2007ns}
\begin{equation}
    F_{ij}=\frac{1}{2} \operatorname{tr}\left[\frac{\partial \textbf{\textsf{C}}}{\partial \theta_i} \textbf{\textsf{C}}^{-1} \frac{\partial \textbf{\textsf{C}}}{\partial \theta_j} \textbf{\textsf{C}}^{-1}\right]
\end{equation}
where $\theta_i$ and $\theta_j$ are from the set of parameters $\boldsymbol{\Theta}$ that influence the covariance and their derivatives are evaluated at the fiducial values $\theta_\text{fid}$, which coincides with the maximum of the likelihood distribution. The Cramer-Rao bound ensures that no unbiased method can measure $\theta_i$ with error bars less than $\sqrt{F_{ii}^{-1}}$. 

For the multipole expansion formalism with an observable like $\delta(\boldsymbol{x})$, the Fisher matrix simplifies to \citep{Taruya:2011tz},
\begin{equation}            
    F_{ij}|_{\ell_\text{max}}=\frac{V_\text{sur}}{4 \pi^2} \sum^{\ell_\text{max}}_{\ell, \ell^{\prime}} \int_{k_{\min }}^{k_{\max }} k^2\text{d}k  \frac{\partial P_{\ell}(k)}{\partial \theta_i}\left[\widetilde{C}^{\ell \ell^{\prime}}(k)\right]^{-1} \frac{\partial P_{\ell^{\prime}}(k)}{\partial \theta_j}\,,
\end{equation}
where the \textit{reduced} covariance is given by 
\begin{equation}\label{eq:Cov_ell}
    \widetilde{C}^{\ell \ell^{\prime}}(k)=(2 \ell+1)(2 \ell^{\prime}+1)\int_0^1 \text{d}\mu \mathcal{L}_{\ell}(\mu) \mathcal{L}_{\ell^{\prime}}(\mu)P_\text{obs}(k,\mu)^2\,,
\end{equation}
which is a matrix of sub-covariance matrices for each multipole combination. Whilst we do not include $k$-mode coupling in these sub-matrices, modelling the covariance between multipoles is necessary for an unbiased analysis \citep{Grieb:2015bia}. We highlight here, how our approach is avoiding any of the damping terms (e.g. due to the beam) entering the Fisher derivatives $\partial P_{\ell}/\partial\theta_i$, which are sensitive only to the shape of the \textit{cosmological} signal coming from \autoref{eq:P_ell} and \autoref{eq:P_HI_sig}. The loss of information from any instrumental or observational damping will be contained in the covariance \eqref{eq:Cov_ell}, through the factor $P_\text{obs}$ (\autoref{eq:P_HI_obs}).

For computing the derivatives for the Fisher matrix, we follow \citet{Euclid:2019clj} and say that given \autoref{eq:P_ell} we have
\begin{equation}
    \frac{\partial P_{\ell}(k)}{\partial \theta_i} = (2\ell+1) \int_{0}^1 \mathrm{~d} \mu \frac{\partial}{\partial \theta_i}\left[P(k,\mu)\mathcal{L}_\ell(\mu)\right]\,,
\end{equation}
since the integration limits do not depend on the parameters, and hence the derivative can be taken inside the integral. We then compute all derivatives numerically using a 5-point stencil, generically defined for a function $\mathcal{F}$ as
\begin{equation}
    \frac{\partial\mathcal{F}}{\partial\theta} = \frac{-\mathcal{F}(\theta+2\delta \theta) + 8\mathcal{F}(\theta+\delta \theta) - 8\mathcal{F}(\theta-\delta \theta) + \mathcal{F}(\theta-2\delta \theta)}{12\, \delta \theta} \,,
\end{equation}
where $\delta\theta$ is chosen such that it gives converged derivatives. Generally, we find that $\delta\theta\txteq{\sim}10^{-2}$ is an appropriate choice for all parameters.

\medskip
\noindent
\textbf{Cross-correlation and multi-tracer}: A cross-correlation power spectrum between the \hi\ and galaxies is given by (dropping redshift dependence for brevity)
\begin{equation}
\begin{split}
    P_{\hi,\text{g}}(k^\text{f}, &\, \mu^\text{f}) = \alpha_{\|}^{-1} \alpha_{\perp}^{-2}\,r_{\hi,\text{g}}\overline{T}_\hi\,\Big(b_\hi + f\mu^2 + b^\hi_\phi f_{\mathrm{NL}} \mathcal{M}(k)^{-1}\Big) \\
    & \times\Big(b_\text{g} + f\mu^2 + b^\text{g}_\phi f_{\mathrm{NL}} \mathcal{M}(k)^{-1}\Big) \, P_{\rm m}(k)
    \,,
\end{split}
\end{equation}
where $r_{\hi,\text{g}}(z)$ is the cross-correlation coefficient between the two tracers to modulate stochastic suppression in the cross-correlation. This is expected to tend to unity on large scales, so it should not be an extra source of degeneracy for \fNL\ constraints. Similarly to the single tracer, the \textit{multi-tracer} Fisher matrix will be given by
\begin{equation}
    F_{ij}|_{\ell_\text{max}}=\frac{V_\text{bin}}{4 \pi^2} \sum^{\ell_\text{max}}_{\ell, \ell^{\prime}} \int_{k_{\min }}^{k_{\max }} k^2\text{d}k  \frac{\partial \boldsymbol{P}_{\tau,\ell}(k)}{\partial \theta_i}\left[\textbf{\textsf{C}}_{\tau\tau^\prime}^{\ell \ell^{\prime}}(k)\right]^{-1} \frac{\partial \boldsymbol{P}_{\tau^\prime,\ell^{\prime}}(k)}{\partial \theta_j}
\end{equation}
but we are now implicitly integrating over the contributions within $\tau\txteq{=}\{\alpha,\beta,\text{X}\}$ i.e.\ the different different generic tracers $\alpha$ and $\beta$, plus their cross-correlation X. For the combination of \hi\ intensity mapping and a galaxy survey, we would have $\alpha\txteq{\equiv}\hi$ and $\beta\txteq{\equiv}\text{g}$. The data vector $\boldsymbol{P}_{\tau,\ell}$ represents the signal-only (``non-observed'') power spectra (i.e.\ without noise and instrumental effects) containing each tracer combination and multipole i.e.
\begin{equation}
    \boldsymbol{P}_{\tau,\ell}(k) = \{\boldsymbol{P}_{\alpha,\ell}(k),\boldsymbol{P}_{\beta,\ell}(k),\boldsymbol{P}_{\text{X},\ell}(k)\} 
\end{equation}
with each element its own $n_\ell \txteq{\times} n_k$ vector. For the multipoles analysis, the covariance is given by
\begin{equation}
    \textbf{\textsf{C}}^{\ell\ell^\prime}_{\tau\tau^\prime}(k)=\left(\begin{array}{ccc}
    C_{\alpha\alpha}^{\ell \ell^{\prime}} & C_{\alpha\beta}^{\ell \ell^{\prime}} & C_{\alpha\text{X}}^{\ell \ell^{\prime}}\\
     & C_{\beta\beta}^{\ell \ell^{\prime}} & C_{\beta\text{X}}^{\ell \ell^{\prime}}\\
     &  & C_{\text{X}\text{X}}^{\ell \ell^{\prime}}
\end{array}\right)\,,
\end{equation}
where each element represents the reduced covariance as before, given by the below;
\begin{equation}
    C_{\alpha\alpha}^{\ell \ell^{\prime}}(k)=(2 \ell+1)(2 \ell^{\prime}+1)\int_0^1 \text{d}\mu \mathcal{L}_{\ell}(\mu) \mathcal{L}_{\ell^{\prime}}(\mu)P^\text{obs}_\alpha(k,\mu)^2\,,
\end{equation}
\begin{equation}
    C_{\alpha\beta}^{\ell \ell^{\prime}}(k)=(2 \ell+1)(2 \ell^{\prime}+1)\int_0^1 \text{d}\mu \mathcal{L}_{\ell}(\mu) \mathcal{L}_{\ell^{\prime}}(\mu)P_\alpha(k,\mu)P_\beta(k,\mu)\,,
\end{equation}
\begin{equation}
    C_{\alpha\text{X}}^{\ell \ell^{\prime}}(k)=(2 \ell+1)(2 \ell^{\prime}+1)\int_0^1 \text{d}\mu \mathcal{L}_{\ell}(\mu) \mathcal{L}_{\ell^{\prime}}(\mu)P^\text{obs}_\alpha(k,\mu)P_\text{X}(k,\mu)\,,
\end{equation}
\begin{equation}
    C_{\text{X}\text{X}}^{\ell \ell^{\prime}}(k)=\frac{(2 \ell+1)(2 \ell^{\prime}+1)}{2}\int_0^1 \text{d}\mu \mathcal{L}_{\ell}(\mu) \mathcal{L}_{\ell^{\prime}}(\mu)\left[P_\text{X}^2(k,\mu) + P^\text{obs}_\alpha(k,\mu)P^\text{obs}_\beta(k,\mu)\right]\,.
\end{equation}
Each element is its own sub-matrix of size $n_\ell\txteq{\times}n_k$. As already mentioned, we assume no mode coupling between $k$, thus the only non-zero off-diagonal components in the matrix come from multipole ($\ell\ell^\prime$) and tracer ($\tau\tau^\prime$) covariance.\newline
\\
For further reading and related works on forecasting in the context of \hi\ intensity mapping, we recommend \citet{Camera:2013kpa,Bull:2014rha,Fonseca:2015laa,Alonso:2015sfa,Alonso:2015uua,Pourtsidou:2016dzn,Obuljen:2017jiy,Bernal:2019jdo,Berti:2023viz}.

\section{Observational effects and cosmological information gain}\label{sec:ObsImpact}

A challenge going forward for intensity mapping analyses will be how to model the influence of observational and instrumental effects, without biasing the inferred cosmological information. In this appendix, we demonstrate how subtle modelling choices can yield different outcomes, using the angular smoothing from the radio telescope beam as an example.

Previous works \cite[e.g.][]{Bull:2014rha,Bull:2015lja,Pourtsidou:2015mia,Pourtsidou:2016dzn,Karagiannis:2022ylq} have sought to separate instrumental effects, such as the beam, from the pure signal, which in principle will provide a cleaner isolation of the cosmological information. This can be achieved with a simple deconvolution of the beam at the power spectrum level, so that \autoref{eq:P_HI_obs} becomes
\begin{equation}\label{eq:deconv_P_HI}
    \tilde{P}^{\rm obs}_\hi(k,\mu) = P_\hi(k,\mu) + \frac{P_{\rm N}}{\mathcal{B}^2_{\rm beam}(k,\mu)}\,,
\end{equation}
then the beam still enters the covariance, but now as an \textit{inflationary} factor through
\begin{equation}\label{eq:deconv_cov}
    \tilde{\sigma}^2(k, \mu)=\frac{\tilde{P}_\text{obs}^2(k,\mu)}{N_{\operatorname{modes}}(k, \mu)}\, = \frac{P_\hi(k,
    \mu) + \frac{P_\text{N}}{\mathcal{B}_\text{beam}^2(k,\mu)}}{N_{\operatorname{modes}}(k, \mu)}\,.
\end{equation}
However, treating these effects as part of the covariance (i.e.\ deconvolving the beam from the measured signal and including its inverse in the noise) is \textit{not equivalent} to including them directly in the signal model of the power spectrum multipoles, which are the realistic measurable from a 21cm intensity mapping survey. Appendix B in \citet{Bernal:2019jdo} demonstrated that when the damping from the beam (or other instrumental effects) is placed entirely in the covariance, the variance of the power spectrum involves terms of the form 
\(\int d\mu\, 1/\mathcal{B}^2(k,\mu)\), 
which diverge as $\mathcal{B}\txteq{\rightarrow}0$, causing covariance to asymptote to infinity for the multipoles. However, earlier works \citep{Seo:2007ns,Wang:2012bx} provide an important complementary warning: damping or smoothing terms, whether from redshift errors, instrumental beams, or non-linear suppressions, represent \textit{losses of information} but can appear to break degeneracies and tighten constraints, if not carefully treated. 

\begin{figure}
    \centering
    \includegraphics[width=1\linewidth]{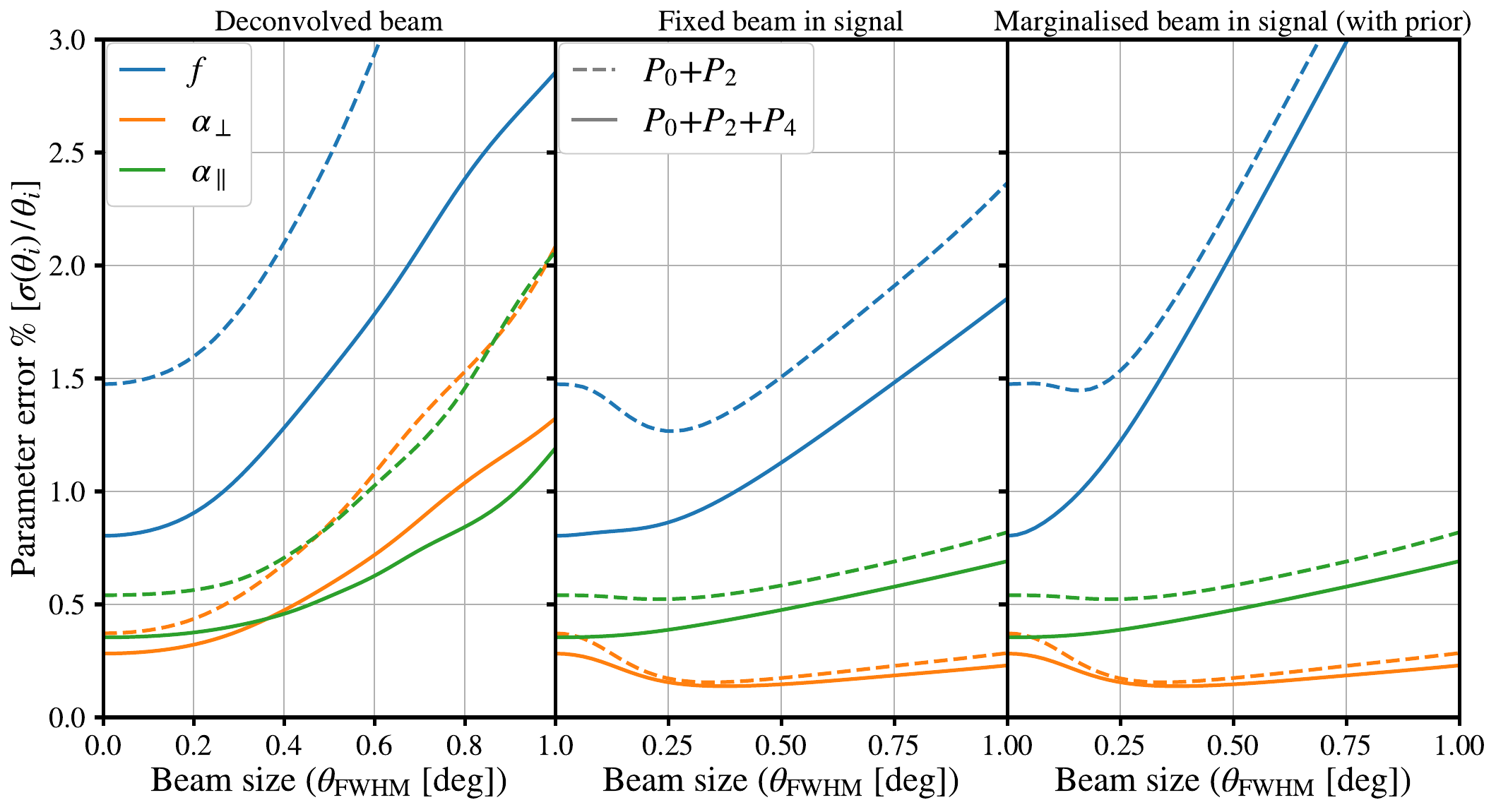}
    \caption{Impact of an increasing telescope beam on the growth rate $f$ and the AP parameters (marginalising over $b_\hi$) for a single redshift bin at $1.0\txteq{<}z\txteq{<}1.2$. The dashed line shows the monopole and quadrupole, with the solid line showing the addition of the hexadecapole. \textit{Left} panel shows the case where the signal is perfectly beam-corrected (deconvolved), moving its impact entirely to the covariance. \textit{Central} panel is the more realistic case where the beam remains in the signal (adopted in this work). \textit{Right} panel again leaves the beam in the signal, but introduces the nuisance parameter $\delta_\text{b}$ to its size (with a tight 1\% precision prior), which we marginalise over. Generally, errors worsen as the beam grows, as expected, but we discuss in the text why the beam can appear to improve constraints in some cases.}
    \label{fig:increasingbeam}
\end{figure}

Our Fisher forecast formalism allowed us to explore this situation, and \autoref{fig:increasingbeam} demonstrates a test where the Gaussian beam width ($\theta_{\rm FWHM}$) was gradually increased from zero while forecasting constraints on cosmological parameters 
$(f, \alpha_\perp, \alpha_\parallel)$. For the case with the deconvolved beam (left panel), i.e. \autoref{eq:deconv_P_HI} and \autoref{eq:deconv_cov}, we see that parameter errors consistently increase with the beam size. To control the integrals in this approach, we had to artificially cut the beam at low values (${<}\,10^{-30}$) to avoid infinity errors. Whilst this appears to provide sensible results, such a deconvolution is extremely challenging for real data because the telescope beam (and other instrumental effects) is not known exactly, plus implementing it on the curved sky further increases the complexity. Since the covariance typically remains fixed in an analysis (e.g. it is not easily recomputed inside an MCMC), uncertainties associated with the beam can not be captured with nuisance parameters if such a deconvolution is attempted, thus there is a high risk of a biased analysis.

The central panel of \autoref{fig:increasingbeam} shows the more realistic situation where the beam remains in the signal, i.e. the original \autoref{eq:P_HI_obs_orig}. Here, we see mild evidence that the beam is enhancing cosmological information gain. This can be understood by considering that the wider beam is creating a more anisotropic field, boosting the SNR of $P_2$ and $P_4$, which will be better at breaking parameter degeneracies, and thus deliver enhanced constraints. This is a limited effect since at some point we should lose information faster than we gain anisotropy. $\alpha_\perp$ improves because when the power is preferentially radial, the derivative of the multipoles with respect to $\alpha_\perp$ changes more sharply than that for $\alpha_\parallel$, thus $\partial P_\ell/\partial\alpha_\perp$ becomes more orthogonal to other derivatives, so its correlation with them decreases, tightening its marginal error. On the other hand, since the beam has little impact on radial modes, any beneficial effect on $\alpha_\parallel$ is smaller, and we even see a mild degradation as the beam damping reduces the usable mode count.

To make the situation more realistic, we introduced a simple beam nuisance parameter, $\delta_{\rm b}$, which perturbs the fiducial beam width such that $\theta_{\rm FWHM}\txteq{\rightarrow}\theta_{\rm FWHM} \txteq{+} \delta_{\rm b}$. This parameter was then varied alongside the cosmological parameters to mimic uncertainty in the true telescope beam. The final panel of \autoref{fig:increasingbeam} presents the results of this test. This has little impact on the AP parameters $(\alpha_\perp,\alpha_\parallel)$, as these depend primarily on geometric distortions that shift scale positions rather than on the overall amplitude anisotropy governed by the beam. However, there is a substantial reduction in information gain on the growth rate $f$. The inferred constraints on $f$ become highly degenerate with the beam size, requiring a tight $1\%$ prior on $\delta_{\rm b}$, i.e. the additive offset is constrained at the 1\% level of the fiducial $\theta_{\rm FWHM}$. We imposed this Gaussian prior of $\sigma_{\delta_\text{b}}\txteq{=}0.01\theta_{\mathrm{FWHM}}$ into the Fisher matrix as $F_{\delta_\text{b}\delta_\text{b}}\txteq{\rightarrow}F_{\delta_\text{b}\delta_\text{b}} \txteq{+}1/\sigma_{\delta_\text{b}}^2$. This degeneracy arises because both the beam width and the growth rate control the degree of radial anisotropy in the observed field: a broader beam or stronger growth both make the power appear more radially elongated. 

So whilst in some cases the anisotropic beam appears to tighten parameter constraints in \autoref{fig:increasingbeam} by breaking degeneracies between anisotropy-sensitive parameters, this information is instrumental rather than cosmological. The telescope beam re-weights angular modes in a known, fixed pattern, which in a Fisher analysis can mimic the effect of additional anisotropic information. In practice, 
apparent improvement in parameter constraints induced by the beam in the Fisher forecasts is not expected to persist in full likelihood analyses. In a realistic MCMC on data, where both the signal and the model include the same beam, smoothing should not generate new cosmological information; it only reweights the available modes. The apparent ‘information gain’ seen in the Fisher matrix thus arises from the altered geometry of the model derivatives rather than from truly increased parameter sensitivity.

In future work, we will explore this further with realistic simulations to validate that any recovered cosmological information is unbiased. As the MeerKAT beam is now being characterised with increasingly high precision  \citep{2023AJ....165...78D}, simulations will also provide the ideal environment to test how accurately the beam must be known and whether dedicated nuisance parameters are required to capture its residual uncertainty. This will ultimately frame the challenge of breaking the growth-rate--beam degeneracy, where we will explore the additional information available in, for example, clustering $\mu$-wedges \citep{Jennings:2015lea,Cunnington:2020mnn} or a \textit{tomographic} analysis; leveraging the differing redshift evolution of the beam and the growth rate.






\end{appendices}


\bibliography{sn-bibliography}

\end{document}